\documentclass[reprint, amsmath,amssymb,groupedaddress,superscriptaddress, prl,longbibliography,floatfix,preprint,onecolumn, 11pt]{revtex4-2}
\usepackage[T1]{fontenc}
\usepackage{graphicx}
\usepackage{dcolumn}
\usepackage{bm}
\usepackage{amsmath}
\usepackage{lmodern}
\usepackage{mathtools}
\usepackage[colorlinks,citecolor=black,linkcolor=black,urlcolor=black]{hyperref}
\usepackage[normalem]{ulem}
\usepackage{upgreek} 
\hypersetup{
    colorlinks,
    linkcolor= [rgb]{0.1804,0.1882,0.5725},
    citecolor=[rgb]{0.1804,0.1882,0.5725},
    urlcolor=[rgb]{0.1804,0.1882,0.5725}
}
\usepackage{chemformula}
\usepackage{comment}

\begin{document}
\title{Electrically tunable spin-orbit coupled photonic lattice in a liquid crystal microcavity}

\author{Marcin\,Muszy\'nski}
\affiliation{Institute of Experimental Physics, Faculty of Physics, University of Warsaw, Poland}
\author{Przemys\l{}aw\, Oliwa}
\affiliation{Institute of Experimental Physics, Faculty of Physics, University of Warsaw, Poland}
\author{Pavel\,Kokhanchik}
\affiliation{Universit\'e Clermont Auvergne, Clermont Auvergne INP, CNRS, Institut Pascal, F-63000 Clermont-Ferrand, France}
\author{Piotr\,Kapu\'sci\'nski}
\affiliation{Institute of Experimental Physics, Faculty of Physics, University of Warsaw, Poland}
\author{Eva\,Oton}
\affiliation{Institute of Applied Physics, Military University of Technology, Warsaw, Poland}
\author{Rafa\l{}\,Mazur}
\author{Przemys\l{}aw\,Morawiak}
\author{Wiktor\,Piecek}
\affiliation{Institute of Applied Physics, Military University of Technology, Warsaw, Poland}
\author{Przemys\l{}aw\,Kula}
\affiliation{Institute of Chemistry, Military University of Technology, Warsaw, Poland}
\author{Witold\,Bardyszewski}
\affiliation{Institute of Theoretical Physics, Faculty of Physics, University of Warsaw, Poland}
\author{Barbara\,Pi\k{e}tka}
\affiliation{Institute of Experimental Physics, Faculty of Physics, University of Warsaw, Poland}
\author{Daniil\,Bobylev}
\affiliation{Universit\'e Clermont Auvergne, Clermont Auvergne INP, CNRS, Institut Pascal, F-63000 Clermont-Ferrand, France}
\author{Dmitry\,Solnyshkov}
\affiliation{Universit\'e Clermont Auvergne, Clermont Auvergne INP, CNRS, Institut Pascal, F-63000 Clermont-Ferrand, France}
\affiliation{Institut Universitaire de France (IUF), 75231 Paris, France}
\author{Guillaume\,Malpuech}
\email{guillaume.malpuech@uca.fr}
\affiliation{Universit\'e Clermont Auvergne, Clermont Auvergne INP, CNRS, Institut Pascal, F-63000 Clermont-Ferrand, France}

\author{Jacek\,Szczytko}
\email{Jacek.Szczytko@fuw.edu.pl}
\affiliation{Institute of Experimental Physics, Faculty of Physics, University of Warsaw, Poland}

\begin{abstract}
We create a one-dimensional photonic crystal with strong polarization dependence and tunable by an applied electric field.
We accomplish this in a planar microcavity by embedding a cholesteric liquid crystal (LC), which spontaneously forms a uniform lying helix (ULH). The applied voltage controls the orientation of the LC molecules and, consequently, the strength of a polarization-dependent periodic potential. It leads to opening or closing of photonic band gaps in the dispersion of the massive photons in the microcavity. In addition, when the ULH structure possesses a molecular tilt, it induces a spin-orbit coupling between the lattice bands of different parity. This interband spin-orbit coupling (ISOC) is analogous to optical activity and can be treated as a synthetic non-Abelian gauge potential.
Finally, we show that doping the LC with dyes allows us to achieve lasing that inherits all the above-mentioned tunable properties of LC microcavity, including dual and circularly-polarized lasing.
\end{abstract}

\maketitle

\section{Introduction}
The physics of microcavities turned out to be extremely rich within the last decades and directly related to the diversity of materials that can be embedded inside. These embedded materials can deeply modify the cavity mode properties, and conversely, the material response can be affected by the cavity, even beyond its optical properties. A large variety of semiconductor materials have been utilized, ranging from inorganic semiconductors \cite{Weisbuch92, Kasprzak2006} to organics \cite{keeling2020bose}, 2D materials \cite{dufferwiel2015exciton}, moiré heterostructures \cite{zhang2021van} or DNA molecules \cite{baaske2014single}. One example is when semiconductor excitons strongly couple to cavity modes, giving rise to hybrid exciton-polariton modes \cite{kavokin2017microcavities, carusotto2013quantum} with mixed light-matter properties. One can also cite quantum Hall materials, which can have their unique topological insulator character broken by the coupling to the cavity modes \cite{appugliese2022breakdown}, and also polariton chemistry, where chemical reactions are modified by the strong coupling of molecules with the confined light modes \cite{garcia2021manipulating}. The essential degrees of freedom that have to be controlled in photonic systems, and, in particular, in microcavities, are the polarization and orbital angular momentum of light. In that perspective, the spin-orbit coupling (SOC) of light appears as a crucial tool \cite{Liberman_(1992)_Phys.Rev.A,Kavokin_(2005)_Phys.Rev.Lett.,Bliokh_(2009)_J.Opt., Bliokh_(2015)_Nat.Photonics, Cardano_(2015)_Nat.Photonics}. In microcavities, an intrinsic light SOC is due to the TE-TM spliting of photonic modes which is at the origin of the optical spin Hall effect \cite{Kavokin_(2005)_Phys.Rev.Lett.,Leyder_(2007)_Nat.Phys.} and of the topologically non-trivial nature of photonic modes in microcavities \cite{gianfrate2020measurement}. Another key ingredient for photonic mode engineering in these systems is the in-plane potential implementing a periodic lattice. The most efficient approach to date is to etch the cavities to realize lattices of micropillars~\cite{jacqmin2014direct,whittaker2018exciton,su2020observation}. Combining topologically non-trivial photonic modes induced by SOC and appropriate periodic potentials led to the observation of photonic topological insulators \cite{Nalitov2015,klembt2018exciton}, topological lasers \cite{solnyshkov2016kibble,st2017lasing}, and lasers carrying non-zero orbital angular momentum \cite{carlon2019optically}.

The materials known for their unique capabilities to control light polarization on demand, and therefore perspective for topological photonics, are liquid crystals (LCs). Their photonic properties have been studied for more than one century, but mostly in the context of light reflection and transmission, which gave rise to various applications such as displays, optical filters, adaptive optical elements, lasers, etc~\cite{ha2008fabrication,mitov2006going}. Recently, a nematic liquid crystal was embedded in a microcavity, which allowed to synthesize a tunable photonic SOC \cite{Rechcinska_Science2019} and to observe a variety of original effects such as photonic persistent spin helix \cite{krol2021realizing, muszynski2022realizing} or Hermitian and non-Hermitian topological transitions \cite{krol2022annihilation}. 
Another LC phase which not explored so far in that context is the cholesteric LC (CLC), which is induced by doping a nematic LC with a chiral dopant. As a result, the LC molecules self-assemble with a twist, producing a helical structure called uniform lying helix (ULH).   The helical axis can be oriented parallel to the substrate in a  structure by imposing a topological frustration. Naturally, a lying helix induces a periodic variation of refractive index. Moreover, CLCs are known to possess a quick response to an applied electric field, through the flexoelectric  effect~\cite{patel1987flexoelectric}.

In this work, we report the fabrication of structures, where a CLC is embedded in a microcavity. We demonstrate that the resulting ULH induces an in-plane periodic potential acting on the photonic modes. It results in the formation of a spin-polarized lattice band structure, reversibly tunable by an external voltage controlling the CLC molecular orientation. We show that when two cavity modes are brought in resonance, photonic SOC efficiently couples the periodic potential bands of different parity and linear polarization. This realizes an analogue of a 1D chain with two components per site, being tunable and showing topological properties similar to the Su-Schrieffer-Heeger chain. Finally, by introducing a dye within the microcavity, we demonstrate the capability of our platform to achieve spin-polarized and dual lasing. 

\section{Results}

\begin{figure}[hbt]
    \centering 
    \includegraphics[width=16 cm]{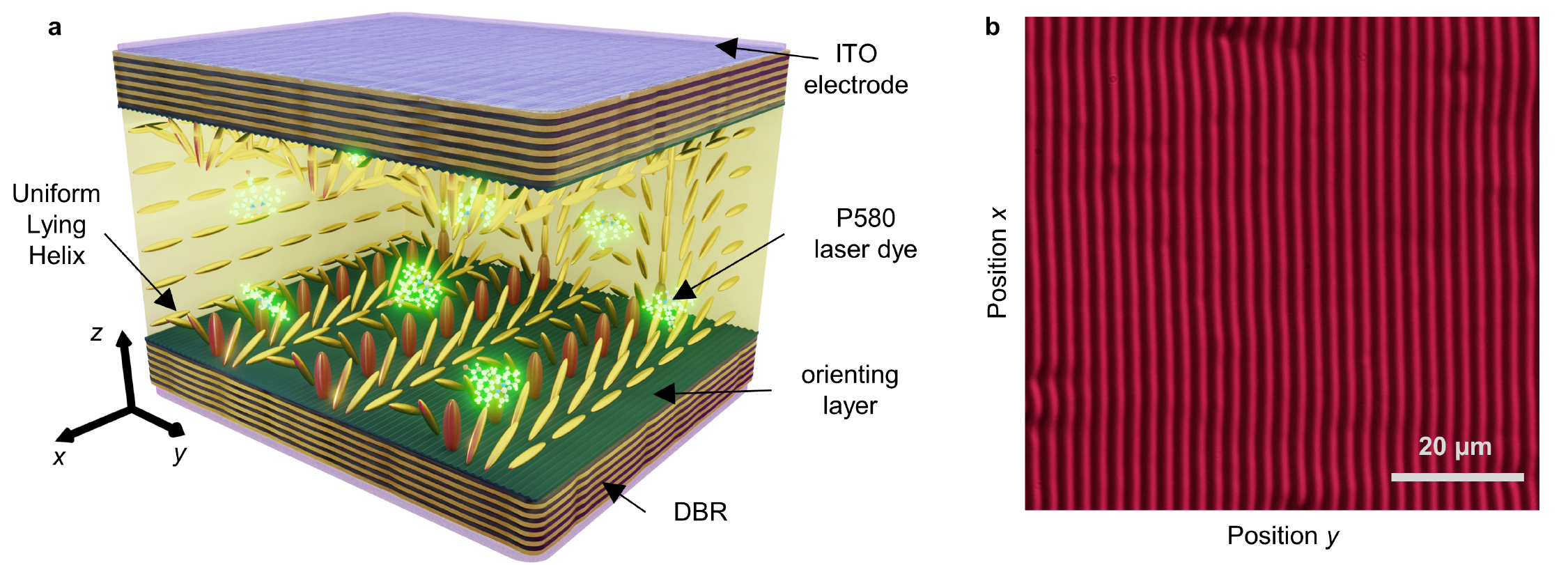}
    \caption{ULH texture inside optical microcavity. (a) Scheme of the dye-doped optical microcavity with an embedded stabilized ULH structure. The rows of molecules whose directors are oriented vertically are colored in red to indicate the period of the lattice. (b) Polarizing optical microscopy image of light transmitted through a ULH texture embedded inside LC microcavity.} 
    \label{im:Fig1}
\end{figure}

The fabrication of our samples depicted in Fig.~\ref{im:Fig1}a is described in the Methods section. The samples are based on an optical microcavity composed of two SiO$_2$/TiO$_2$ distributed Bragg reflectors (DBRs), covered with transparent electrodes of indium tin oxide (ITO). 
The cavity is filled with a nematic LC and a chiral dopant, which induces a helical twist in the LC matrix, producing a supramolecular helical structure with a helical pitch (double period, $p$) of the order of several micrometers~\cite{Oton2017,bisoyi2018stimuli}. By treating the orienting layer with homeotropic anchoring, the surfaces impose a topological frustration that induces a ULH, with the helical axes parallel to the substrate. Cross-polarized transmission microscopy shows a variable periodic refractive index (Fig.~\ref{im:Fig1}b) confirming the ULH configuration~\cite{Komitov1999}.
The variety of possible applications of the proposed system increases with the integration of light emitters, so we doped the system with the organic laser dye pyrromethene 580 (P580), which, when externally excited, can transit the system to a nonlinear regime with emitted polarized laser light.

The proposed optical platform can be fabricated using a range of different liquid crystal mixtures. To demonstrate the versatility of our approach and various ULH-induced effects, we prepared two samples. Sample A, with a higher birefringence of the CLC mixture, was used in the first part of the work to demonstrate band structure tunability by an external electric field. Sample B has a CLC mixture with lower birefringence but exhibits a deeper photonic potential thanks to the higher homogeneity along the cavity axis $z$.
It was used in the second part for generating a 1D lattice in the presence of SOC and for demonstrating lasing.
 
\begin{figure}[hbt]
    \centering 
    \includegraphics[width=16 cm]{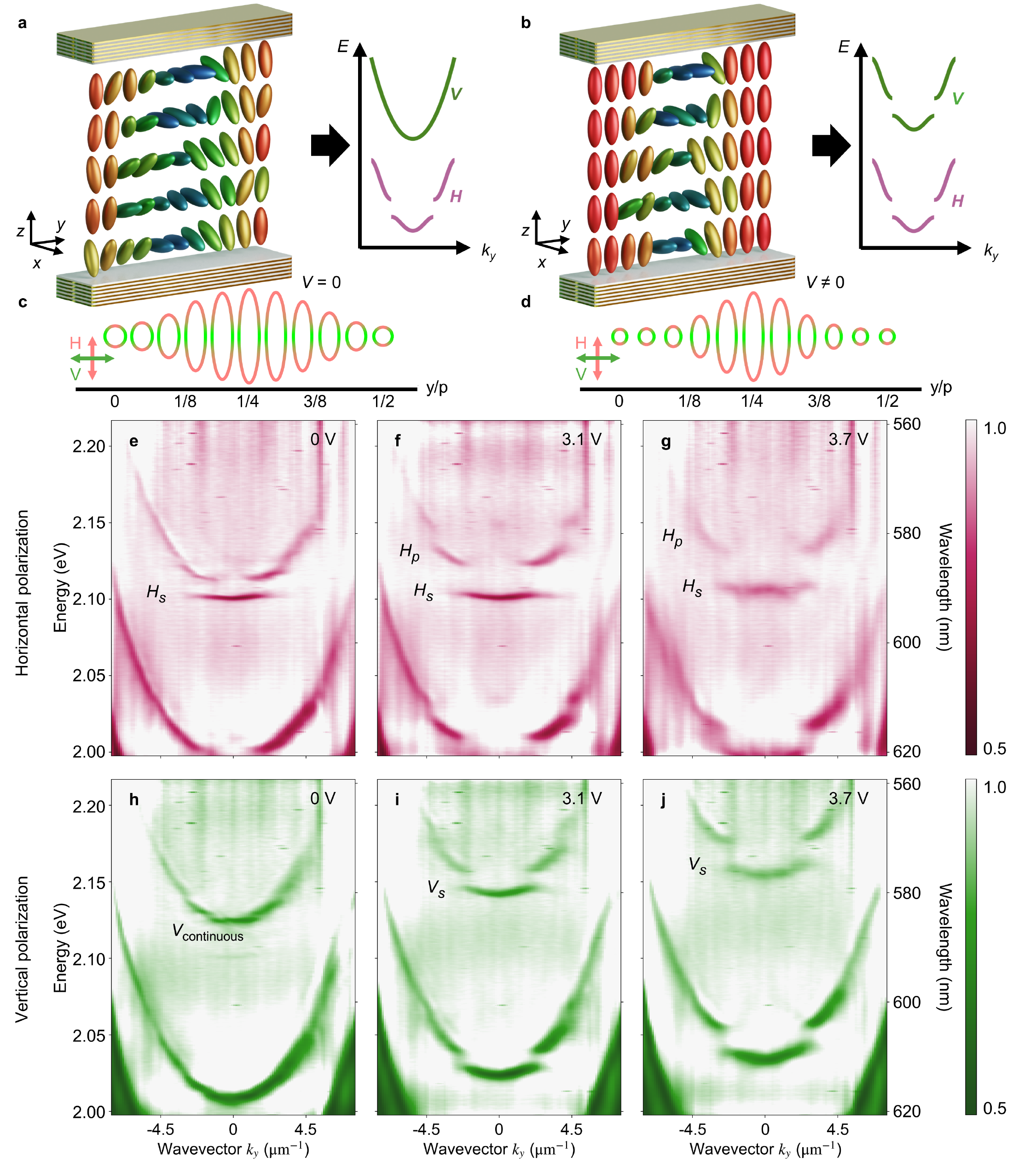}
    \caption{The tunable band structure. Illustrative representation of the LC director arrangement inside the microcavity and the corresponding schematic dispersion relations of the horizontal ($H$, pink) and vertical ($V$, green) modes for (a) zero and (b) non-zero voltage. Effective refractive index ellipses for $H$ and $V$ polarizations showing periodical dependence along the ULH axis $y$ over half pitch $p$ and obtained by averaging director distributions shown in (a) and (b) along microcavity axis $z$ for (c) zero and (d) non-zero voltage, respectively. Reflectivity spectra of (e-g) horizontally and (h-j) vertically polarized band structures measured for different voltages (indicated in top right corners) applied to the sample A.}
    \label{im:Fig2}
\end{figure}

First, we study sample A. An optical cavity transforms photons into 2D massive quasiparticles \cite{Kavokin2003}. These quasiparticles experience a potential defined by the effective refractive index obtained by averaging along the microcavity axis $z$. This averaging can be easily performed for a known LC director distribution inside the microcavity.
The director distribution in the absence of an external voltage naturally consists of a perfect ULH aligned with the $y$ axis and superimposed director fluctuations from the perfect ULH-defined orientation (Fig.~\ref{im:Fig2}a)~\cite{martinand1972electric}. The effective refractive index distribution obtained by averaging along $z$ for both polarizations is shown in Fig.~\ref{im:Fig2}c in the form of effective refractive index ellipses. Remarkably, only $H$ polarization (pink, along $x$ axis) has a periodic variation of refractive index (potential), while $V$ polarization has a constant refractive index (green, along $y$ axis). This should result in gapped $H$-polarized dispersion and continuous $V$-polarized dispersion, as shown in the right part of Fig.~\ref{im:Fig2}a. Under a non-zero applied voltage, the molecules inside the microcavity are reoriented according to the so-called dielectric effect~\cite{rudquist1994linear}. It induces a gradual increase of the refractive index contrast for both polarizations with growing voltage, thus deepening both potentials, opening a gap for $V$ polarization, and increasing the already existing gap for $H$ polarization (Fig.~\ref{im:Fig2}d). For more details, see Section~I.A of~\cite{suppl}.

Our experiment based on angular-dependent reflectivity measurements (see Methods and Section~II in~\cite{suppl}) confirms these expectations. Without any voltage applied to the sample (Fig.~\ref{im:Fig2}e,h), the bandgap is visible only in the $H$ polarization, while the spectrum remains gapless (within the spectrometer resolution and line broadening) in the $V$ polarization. The so-called $s$-band of lowest energy is observed in the $H$ polarization (denoted $H_s$, $E\approx 2.1$~eV).
Under non-zero external voltage, an energy gap emerges in the $V$ polarization as well, as shown in Fig.~\ref{im:Fig2}i,j, opening from 0~meV at 0.0~V to 20~meV at 3.7~V. The whole band structure in $V$ polarization is shifted towards higher energies.
The process of gap opening is reversible, as the liquid crystal director fluctuations recover when the applied voltage is turned off. We also observed the increase of the gap for $H$ polarization, from 12~meV at 0.0~V to 25~meV at 3.7~V (Fig.~\ref{im:Fig2}f,g). Beside the symmetric $s$-band, another gap opens in $H$ polarization for higher energies (around 2.14~eV), thus forming an additional antisymmetric $p$-band (denoted $H_p$), as it is known in solid state physics and photonic structures~\cite{st2017lasing}. 

\begin{figure}[hbt]
    \centering 
    \includegraphics[width=14.5 cm]{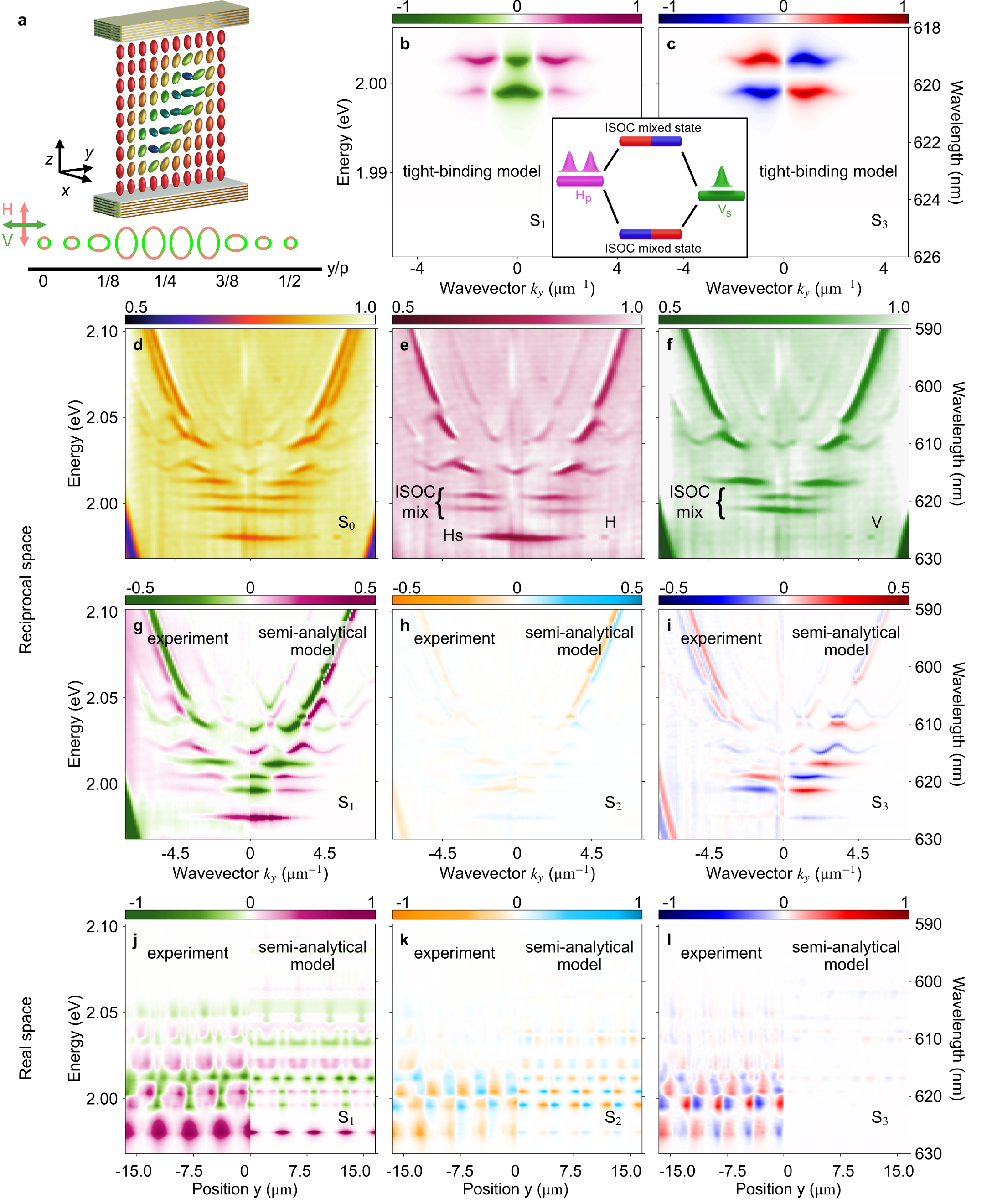}
    \caption{Spin-orbit coupling in ULH LC microcavity. (a) (top) Illustration of the LC director distribution inside the microcavity and (bottom) effective refractive index ellipses. $sp$-hybridized bands obtained from the tight-binding Hamiltonian Eq.~\eqref{tb_Hamiltonian} with colors corresponding to (b) $S_1$ and (c) $S_3$ Stokes parameters ($S_2=0$ everywhere); the inset shows a scheme of hybridization. The $k$-space reflectivity spectra measured for (d) total intensity ($S_0$ Stokes parameter) and orthogonal linear (e) horizontal $H$ and (f) vertical $V$ polarization of light. The degree of polarization is expressed by Stokes parameters $S_1$, $S_2$, and $S_3$ in (g,h,i) reciprocal and (j,k,l) real space, respectively. The left panels of the Stokes parameter maps correspond to the experimental data, while the right panels represent the semi-analytical model (see Section~V in~\cite{suppl}). The ISOC-mixed eigenstates appear near 2.0~eV as a result of coupling between $p$-band of $H$ polarization and $s$-band of $V$ polarization.}
    \label{im:Fig3}
\end{figure}

Therefore, we have demonstrated that sample A shows an on-demand reconfigurability of the band structure. It is important to stress that the strong optical activity (OA) of a CLC helix~\cite{de1993physics} is suppressed for sample A because of the mirrors, restoring the inversion symmetry. Consequently, the $H$ and $V$ modes are uncoupled. However, this is not always the case as we show below. If the inversion symmetry is broken, then the linear polarizations should be coupled by an anti-symmetric SOC term, in the simplest case, linear in the wavevector. In this general case, an effective Hamiltonian describing the cavity modes of two polarizations $H$ and $V$ in the polarization-dependent periodic potential created by the ULH on the ($H$,$V$) basis in real space reads:
\begin{widetext}
\begin{equation}
    H_{cont} =
    \begin{pmatrix}
        -\frac{\hbar^2}{2m} \frac{\partial^2}{\partial y^2} + \frac{A_H}{2} [1 - \cos{\left(2qy\right)}] - \frac{\delta}{2} & -2i\alpha \frac{\partial}{\partial y} \\
        -2i\alpha \frac{\partial}{\partial y} & -\frac{\hbar^2}{2m} \frac{\partial^2}{\partial y^2} + \frac{A_V}{2} [1 - \cos{\left(2qy\right)}] + \frac{\delta}{2}
    \end{pmatrix},
    \label{SM_continuous_Hamiltonian}
\end{equation}
\end{widetext}
with $\hbar$ the reduced Planck constant, $m$ the mass of cavity photon, $\delta$ the linear birefringence, $q = 2\pi/p$, and $p$ the pitch of the ULH. $A_{H,V}$ are the potential amplitudes for $H$ and $V$ polarizations. As discussed for sample A, $A_{V}$ can be switched on and off, and $A_{H}$ can be modified by applying a voltage.
The coefficient $\alpha$ describes the SOC resulting from the OA, which is zero for sample A, as explained above. This coupling term is called the Rashba-Dresselhaus SOC (RDSOC), since it appears in the case of balanced Rashba and Dresselhaus SOC contributions in electronics. The physics offered by this term has already been studied in the context of birefringent microcavities~\cite{Rechcinska_Science2019, Ren2021}, although the microscopic origin of the term was different.

In order to combine the effects of the periodic potential and the SOC, we prepared sample B with a different CLC material. The origin of SOC for sample B lies in the particular texture of the director distribution inside the microcavity at zero voltage (Fig.~\ref{im:Fig3}a, top), different from sample A. This distribution combines the homeotropic (vertical) anchoring of the molecules at the mirrors, perfect ULH structure, and a tilt of the molecules around the $x$ axis (the role of director fluctuations is not crucial for sample B). Both straight and tilted director distributions of samples A and B, respectively, are stable. The particular choice of one of them depends on the used LC, deposition techniques, anchoring strength, etc~\cite{jia2021engineering}. Because of the presence of the tilt, the inversion symmetry of the system is broken, and the SOC, suppressed for a straight molecular orientation of sample A, is restored in sample B (see section~I.B of~\cite{suppl} for details). The necessity of the tilt for SOC appearance is additionally confirmed by the exact solution of Maxwell's equations based on the Finite Element Method (see Section~III of~\cite{suppl}). In sample B, both polarizations have a non-zero effective refractive index contrast ($A_{H,V}\neq 0)$ even at zero voltage (Fig.~\ref{im:Fig3}a, bottom), which opens multiple gaps in their dispersions.

In that regime, it is possible to go further and derive from the continuous Hamiltonian \eqref{SM_continuous_Hamiltonian} a tight-binding Hamiltonian describing the SOC between the orthogonally-polarized bands of the periodic potential, as detailed in Section~IV of~\cite{suppl}. Indeed, in the absence of the SOC, each polarization exhibits a band structure formed from its own ladder of states: $s$, $p$, $d$, and so on. The SOC term is an odd function of $k$, and so it couples only the localized states of opposite parity. For instance, the $s$-band in $V$ polarization ($V_s$) and the $p$-band in $H$ polarization $H_p$ are coupled by the SOC, giving rise to what can be called an interband SOC (ISOC). It is worth noting that this contrasts with earlier works on SOC in birefringent cavities, in which the opposite parity of states was satisfied by coupling even and odd $H$ and $V$ modes~\cite{Rechcinska_Science2019, Ren2021}. Here, the $H$ and $V$ modes have the same parity, therefore the even subband $s$ is coupled to the odd subband $p$.
The corresponding tight-binding Hamiltonian reads:
\begin{equation}
    H = 
    \begin{pmatrix}
        E_{H_p} - 2t_{pp} \cos{(k_y p/2)} & -i\alpha \left[ t_{sp,0} - 2 t_{sp,1} \cos{(k_y p/2)} \right] \\
        c.c. & E_{V_s} - 2t_{ss} \cos{(k_y p/2)}
    \end{pmatrix},
    \label{tb_Hamiltonian}
\end{equation}
where $E_{H_p}$ and $E_{V_s}$ are the onsite energies of $H_p$ and $V_s$, respectively, $t_{pp}$ and $t_{ss}$ are the nearest-neighbor couplings between $p$ states of $H$ and $s$ states of $V$, respectively, $t_{sp,0}$ and $t_{sp,1}$ are the onsite and nearest-neighbor couplings between $H_p$ and $V_s$ (ISOC terms), respectively. All couplings are calculated from the Hamiltonian~\eqref{SM_continuous_Hamiltonian} (including the ISOC terms) using the standard perturbation theory on the basis of $s$ and $p$ states of a harmonic oscillator. Explicit expressions are provided in Section~IV of~\cite{suppl}. In the limit of strongly localized modes ($\alpha t_{sp,1},\ t_{ss},\ t_{pp} \ll \alpha t_{sp,0}$), the spectrum of the Hamiltonian~(\ref{tb_Hamiltonian}) is shown in Fig.~\ref{im:Fig3}b,c, with color representing the Stokes parameters $S_1$ and $S_3$.

We performed the experiment for sample B in order to confirm our predictions (Fig.~\ref{im:Fig3}d-l, left panels). Fig.~\ref{im:Fig3}d-f present the angle-resolved reflectivity spectra measured for total intensity (Stokes parameter $S_0$), horizontal $H$, and vertical $V$ polarizations of detected light, respectively. The right parts of the panels 
are obtained by a model beyond the tight-binding limit based on a semi-analytical solution of Maxwell's equations~\cite{Oliwa_(2024)_Phys.rev.res.} using the coefficients of the Fourier decomposition of the dielectric tensor of the material inside the cavity as fitting parameters (details are given in Section~V of~\cite{suppl}. 

The presence of a deeper photonic potential, compared to Fig.~\ref{im:Fig2}, leads to the opening of a greater number of photonic bandgaps, as well as a deeper localization of light: a large effective mass of the trapped photons makes the lowest bands almost flat. 
In a system with uncoupled $H$ and $V$ polarizations, each of them should possess a band structure with its own ladder of states. However, we observe that the states around 2~eV, expected to be $H_p$ and $V_s$, show up in both polarizations. Fig.~\ref{im:Fig3}g,h,i show the corresponding Stokes parameters $S_1$, $S_2$, and $S_3$, measured in the reflectivity in $k$-space. It is evident that each of these two coupled states contains both $H$ and $V$ contributions. Furthermore, in the circular polarization degree $S_3$ in Fig.~\ref{im:Fig3}i, these states form a check pattern which is extremely well reproduced by the tight-binding model (Fig~\ref{im:Fig3}b,c) and the semi-analytical model (right part of the panel). These features are also well reproduced by the exact numerical solution of Maxwell's equations with a ULH dielectric tensor including molecular tilt (Section~III of~\cite{suppl}).
In addition, Fig.~S3 of~\cite{suppl} shows the influence of the detuning between the bands $H_p$ and $V_s$ on the degree of circular polarization $S_3$ measured in $k$-space.

Stokes parameters of spatially-resolved ($r$-space) transmission spectrum are shown in Fig.~\ref{im:Fig3}j-l. The lowest energy states in Fig.~\ref{im:Fig3}j portray the spatial periodicity of the refractive index corresponding to the half-pitch of ULH. The check pattern similar to that in Fig.~\ref{im:Fig3}i is visible in the Stokes parameter $S_2$ (Fig.~\ref{im:Fig3}k) for $sp$-hybridized states at energies around 2~eV, which further proves the presence of the ISOC in the system.

\begin{figure}[hbt]
    \centering 
    \includegraphics[width=8 cm]{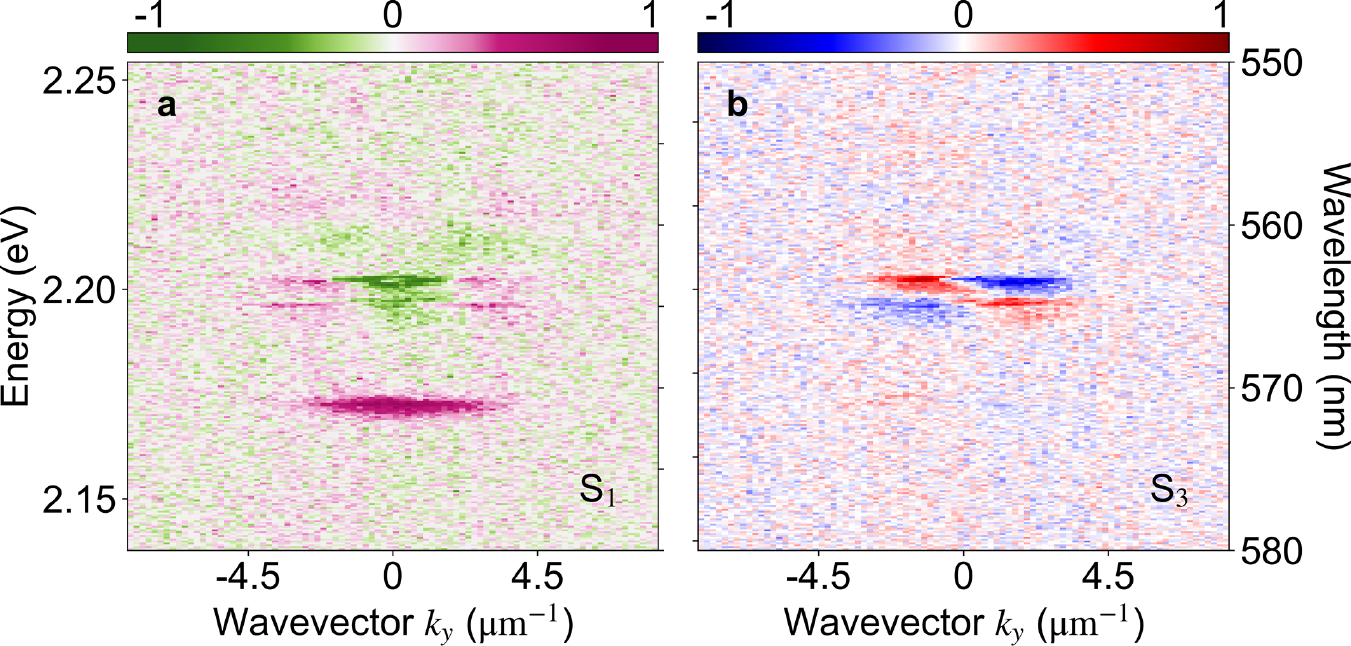}
    \caption{Polarization conserving lasing from ULH LC microcavity. Stokes parameters (a) $S_1$ and (b) $S_3$ of angle-resolved emission spectra collected above the lasing threshold for 0.36 \textmu J/pulse pumping energy.} 
    \label{im:Fig4}
\end{figure}

As an ultimate demonstration of the viability and usefulness of our design and the possibilities that it offers, we harnessed the spin-polarized bands to manipulate the polarization of laser emission. For an optically-pumped dye-doped nematic LC microcavity, lasing takes place at the minimum energy of the $H$-polarized mode, which better overlaps spectrally with the dye gain curve~\cite{muszynski2022realizing}.  
Due to the presence of the photonic potential and the coupling of the $H_p$ and $V_s$ modes, for the ULH microcavity, the energy transfer between the split states provides lasing in both orthogonal linear polarizations of light. Fig.~\ref{im:Fig4}a shows the Stokes parameter $S_1$ of angle-resolved emission spectrum above the lasing threshold for pumping energy of 0.36~\textmu J/pulse (see Fig.~S4 in Section~II of~\cite{suppl} for photoluminescence intensity and linewidth as a function of pumping power). Our experiments demonstrate that lasing occurs not only for the ground state $H_s$, but also for the ISOC mixed states, preserving the polarization structure of the bands. In this regime, the characteristic check pattern in the circular polarization degree $S_3$ is inherited (Fig.~\ref{im:Fig4}b). 

\section{Discussion}

The effective $2\times 2$ Hamiltonian~\eqref{SM_continuous_Hamiltonian} describing the interaction of bands in the continuous limit has a particularly interesting interpretation in terms of a non-Abelian gauge potential~\cite{Rechcinska_Science2019} (similar to the Yang-Mills gauge potential \cite{jin2006,Tokatly2008}), described by the RDSOC constant $\alpha$. Its non-Abelian nature allows it to produce effects, such as Zitterbewegung \cite{Polimeno2021}, even for a constant gauge potential. The ULH, which is known to exhibit various topological defects \cite{wu2022hopfions} characterized by the change of the RDSOC magnitude and orientation, offers even broader opportunities: the variation of RDSOC (corresponding to a non-zero gauge field) can be expected to lead to spin-polarized photonic Landau levels \cite{schine2016synthetic} exhibiting non-zero orbital angular momentum, localized at the topological defects of the LC texture. The change of sign of the RDSOC has already been observed in the present sample (not shown), and the observation of the Landau levels will be a subject for future works.

Another important outlook of our work is related to the topological properties of the tight-binding Hamiltonian~\eqref{tb_Hamiltonian}: similar to the tight-binding Hamiltonian of the Su-Schrieffer-Heeger model~\cite{Su1979}, it can exhibit a topological transition with a change of the winding number~\cite{Delplace2011} when $t_{sp,0}=2t_{sp,1}$, corresponding to the formation of the topological edge states for a finite chain. While such transition is not observed for the $s$ and $p$ bands in our sample, it might potentially be observed for different bands having another tunneling coefficient ratio. In general, the observation of RDSOC in a 1D lattice opens many exciting possibilities, such as the control of the amplitude and phase of the tunneling coefficients~\cite{Kokhanchik2022} and, combined with polarization-dependent losses and gain, it could allow the observation of exceptional points~\cite{krol2022annihilation} and of the non-Hermitian skin effect~\cite{Kokhanchik2023}.

Finally, the results obtained under strong pumping (Fig.~\ref{im:Fig4}) demonstrate that our ULH microcavity can serve as a device with simultaneous laser emission of linearly and circularly polarized non-degenerated lines (dual lasing \cite{zhang2007orthogonally}).

\section{Conclusions}
We have demonstrated a versatile photonic platform based on a cholesteric liquid crystal optical microcavity showing an embedded self-organizing photonic crystal potential. The simple and efficient manufacturing process allows us to obtain, on a macroscopic scale, an electrically-tunable one-dimensional periodic lattice with spin-orbit coupling. In the absence of SOC, the resulting bands are linearly polarized and can be dynamically tuned by the external electric field. The SOC present in the system causes the bands of the periodic lattice of different polarization and parity to couple, leading to the hybridization of states. Our results are well described by the solution of Maxwell's equations, but also by a comprehensive effective Hamiltonian. We utilized ISOC-hybridized states to create a spin-polarized (circularly-polarized) laser operating at room temperature.
This work paves the way for studying the combined effects of SOC, or, more generally, non-Abelian gauge fields, and periodic lattices, including possible topological transitions, with practical applications that include dual spin-polarized lasing.

\section{Methods}

\subsection{Sample preparation}

Both samples (A and B) contained the DBRs centered at 550 nm with 6 pairs of SiO$_2$/TiO$_2$ layers, which were deposited onto commercially available ultra-flat polished glass substrates for STN technology with transparent ITO electrodes. The two DBR substrates were assembled together in a sandwich-like cell, separated by calibrated silica spacers ranging between 2~$\upmu$m and 3.5~$\upmu$m and creating a sample wedge. An orienting polymer layer with homeotropic anchoring was used to align the liquid crystal. The liquid crystal mixtures were prepared with a nematic base, a chiral dopant and a laser dye dopant: Sample A (Fig.~\ref{im:Fig2}) contained a highly birefringent liquid crystal mixture W-2091 ($n_o = 1.565$, $n_e = 1.907$, $\Delta n = 0.342$ at $\lambda = 600 \,\mathrm{nm}$) with a chiral dopant at a concentration of 0.65\%, while sample B (Fig.~\ref{im:Fig3}-\ref{im:Fig4}) contained liquid crystal mixture E7 ($n_o = 1.522$, $n_e = 1.737$, $\Delta n = 0.215$ at $\lambda = 600 \,\mathrm{nm}$) and a chiral dopant concentration of 0.70\%. The chiral dopant used was R-5011 (Daken Chemical) and both mixtures were doped with organic dye P580 at a concentration of 1\%. Producing a stable ULH structure requires precise control of the confinement ratio. The confinement ratio is the relationship between the cavity thickness $d$ and helical pitch $p$: $C=d/p$, and it was kept in a range of $C= 1.85 - 2.75$. This factor is crucial for obtaining a periodic structure with a well-defined alignment and a pitch as small as possible, and it allowed the formation of uniform ULH structures with a decreased number of dislocations. By controlling the chiral dopant concentration, we are able to adjust the ULH pitch.

\subsection{Experimental details}

The optical measurements are performed at room temperature. The scheme of the experimental setup is shown in Fig.~S2 (Section~II of~\cite{suppl}). In reflectivity measurements, the microscope objective with 60x magnification, numerical aperture NA = 0.7, and coverglass correction ring is used for both the excitation and the collection of light. The spot has a diameter of around 10~$\upmu$m. In transmission and photoluminescence measurements, an additional microscope objective with 5x magnification and NA = 0.15 is used for excitation. The set of a quarter-wave plate, a half-wave plate, and a linear polarizer is used to collect light in the selected polarization. For emission measurements, the sample is excited by a $Q$-switched diode-pumped laser with a shot-on-demand experiment (532-nm center wavelength, 2-ns pulse duration). 
To tune the dispersion relation of the cavity modes, the LC microcavity is addressed by an AC waveform generator with a 100 Hz square signal and varying amplitude of 0~--~5~V.

\section*{Acknowledgments}

This work was supported by the National Science Centre grants 2019/35/B/ST3/04147, 2019/33/B/ST5/02658 and 2022/47/B/ST3/02411, 2023/51/B/ST3/03025, and the Ministry of National Defense Republic of Poland Program -- Research Grant MUT Project 13-995 and MUT University grant (UGB) for the Laboratory of Crystals Physics and Technology for the year 2021 and the European Union’s Horizon 2020 program, through a FET Open research and innovation action under the grant agreements No. 964770 (TopoLight). Additional support was provided by the ANR Labex GaNext (ANR-11-LABX-0014), the ANR program "Investissements d'Avenir" through the IDEX-ISITE initiative 16-IDEX-0001 (CAP 20-25), and the ANR project "NEWAVE" (ANR-21-CE24-0019).

\end{document}


\title{Electrically tunable spin-orbit coupled photonic lattice in a liquid crystal microcavity -- Supplementary Information}

\author{Marcin\,Muszy\'nski}
\affiliation{Institute of Experimental Physics, Faculty of Physics, University of Warsaw, Poland}
\author{Przemys\l{}aw\, Oliwa}
\affiliation{Institute of Experimental Physics, Faculty of Physics, University of Warsaw, Poland}
\author{Pavel\,Kokhanchik}
\affiliation{Universit\'e Clermont Auvergne, Clermont Auvergne INP, CNRS, Institut Pascal, F-63000 Clermont-Ferrand, France}
\author{Piotr\,Kapu\'sci\'nski}
\affiliation{Institute of Experimental Physics, Faculty of Physics, University of Warsaw, Poland}
\author{Eva\,Oton}
\affiliation{Institute of Applied Physics, Military University of Technology, Warsaw, Poland}
\author{Rafa\l{}\,Mazur}
\author{Przemys\l{}aw\,Morawiak}
\author{Wiktor\,Piecek}
\affiliation{Institute of Applied Physics, Military University of Technology, Warsaw, Poland}
\author{Przemys\l{}aw\,Kula}
\affiliation{Institute of Chemistry, Military University of Technology, Warsaw, Poland}
\author{Witold\,Bardyszewski}
\affiliation{Institute of Theoretical Physics, Faculty of Physics, University of Warsaw, Poland}
\author{Barbara\,Pi\k{e}tka}
\affiliation{Institute of Experimental Physics, Faculty of Physics, University of Warsaw, Poland}
\author{Daniil\,Bobylev}
\affiliation{Universit\'e Clermont Auvergne, Clermont Auvergne INP, CNRS, Institut Pascal, F-63000 Clermont-Ferrand, France}
\author{Dmitry\,Solnyshkov}
\affiliation{Universit\'e Clermont Auvergne, Clermont Auvergne INP, CNRS, Institut Pascal, F-63000 Clermont-Ferrand, France}
\affiliation{Institut Universitaire de France (IUF), 75231 Paris, France}
\author{Guillaume\,Malpuech}
\affiliation{Universit\'e Clermont Auvergne, Clermont Auvergne INP, CNRS, Institut Pascal, F-63000 Clermont-Ferrand, France}

\author{Jacek\,Szczytko}
\affiliation{Institute of Experimental Physics, Faculty of Physics, University of Warsaw, Poland}
\maketitle

\section{Band structure explained through director distribution}

\subsection{Sample A}

\begin{figure}[hbt]
    \centering 
    \includegraphics[width=\linewidth]{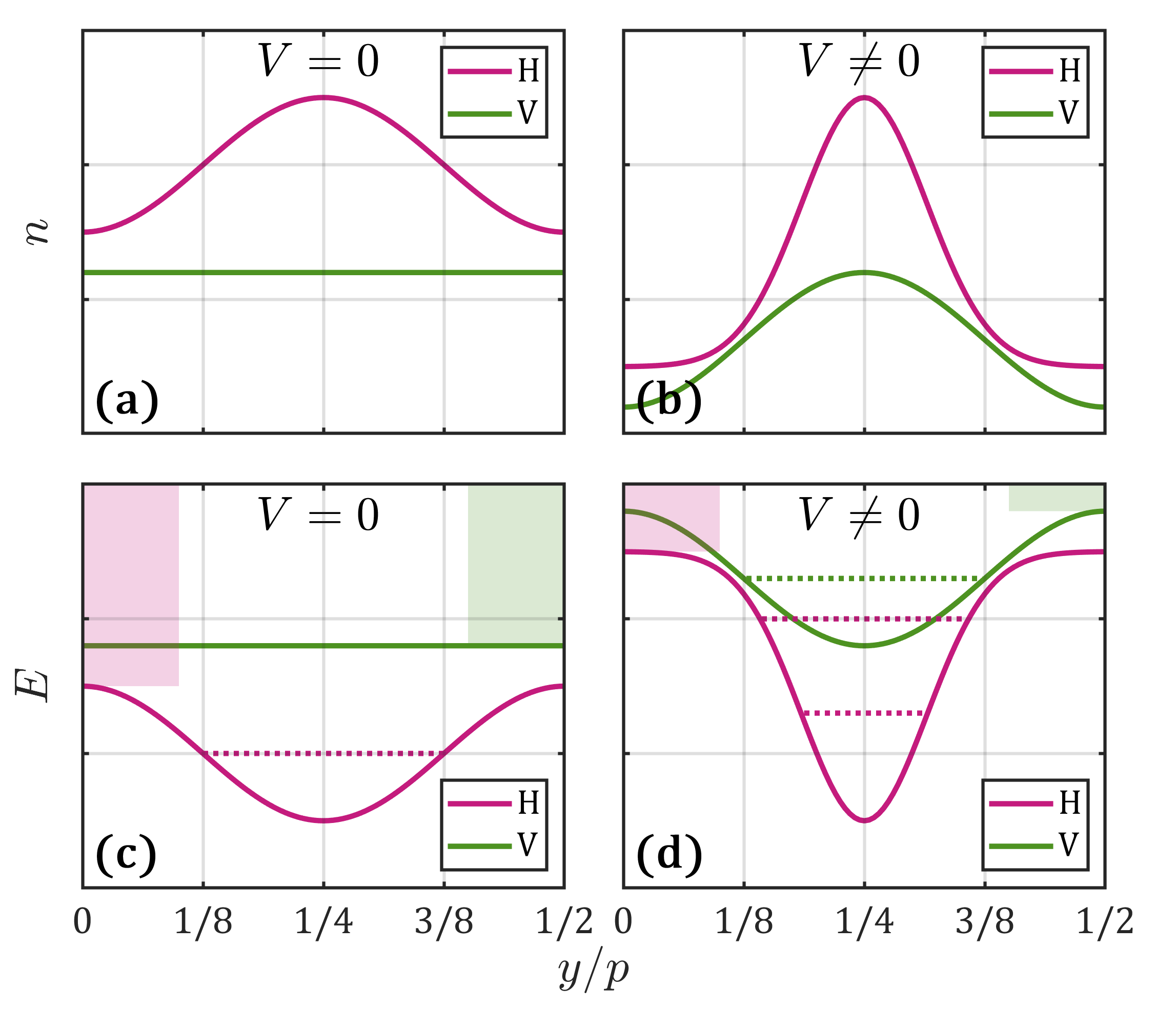}
    \caption{Schematic dependence of refractive indices and corresponding potentials for sample A. Refractive index real space dependence for (a) zero and (b) non-zero voltage. Resulting periodic potentials for (c) zero and (d) non-zero voltage. Pink and green colors correspond to $H$ and $V$ polarizations, respectively. Dotted lines show localized states and colored rectangles show the energy range of the continuum.} 
    \label{im:sample_a}
\end{figure}

The dynamics of the dispersion under an applied electric field can be carefully explained using the director distributions shown in Fig.~2a,b of the main text. The ideal helix creates a perfectly uniform distribution of the director along the microcavity axis $z$. Its projection on microcavity plane $x$-$y$ shows a periodic variation of refractive index along $y$ for one polarization ($H$) and the absence of variation for another polarization ($V$). In reality, the helix is not ideal, and the angular fluctuations~\cite{martinand1972electric} of molecules from ideal orientation occur. Under zero electric field, these deviations are randomly distributed and do not have any preferential direction. In general, their effect is a reduction of the effective refractive index contrast. In other words, if we now average the LC director distribution along $z$ taking into account these fluctuations (Fig.~2a of the main text), the maximum (minimum) refractive index is reduced (increased) for both polarizations in comparison to the ideal helix. The result of this averaging is shown in Fig.~2c of the main text in the form of effective refractive index ellipses for $H$ and $V$ polarizations. It is also shown in a more standard way in Fig.~\ref{im:sample_a}a, as two separate functions, $n_H$ (pink) and $n_V$ (green), representing the effective refractive indices for $H$ and $V$ polarization, respectively. The polarization-dependent potentials created by these refractive index distributions are calculated as:
\begin{equation}
\label{refr_index_to_potential}
V_{H,V} = \frac{\pi \hbar c N}{n_{H,V} L} \propto \frac{1}{n_{H,V}},
\end{equation}
with reduced Planck constant $\hbar$, speed of light $c$, microcavity mode number $N$, and microcavity length $L$. These potentials are schematically depicted in Fig.~\ref{im:sample_a}c. The periodic potential with half-pitch period $p/2$ is present for $H$ and absent for $V$. Naturally, if the potential for $H$ is deep enough, it contains a localized state (shown as a dotted line in Fig.~\ref{im:sample_a}c), exactly as in the experiment (Fig.~2e of the main text).

\begin{figure*}
    \centering 
    \includegraphics[width = 16 cm]{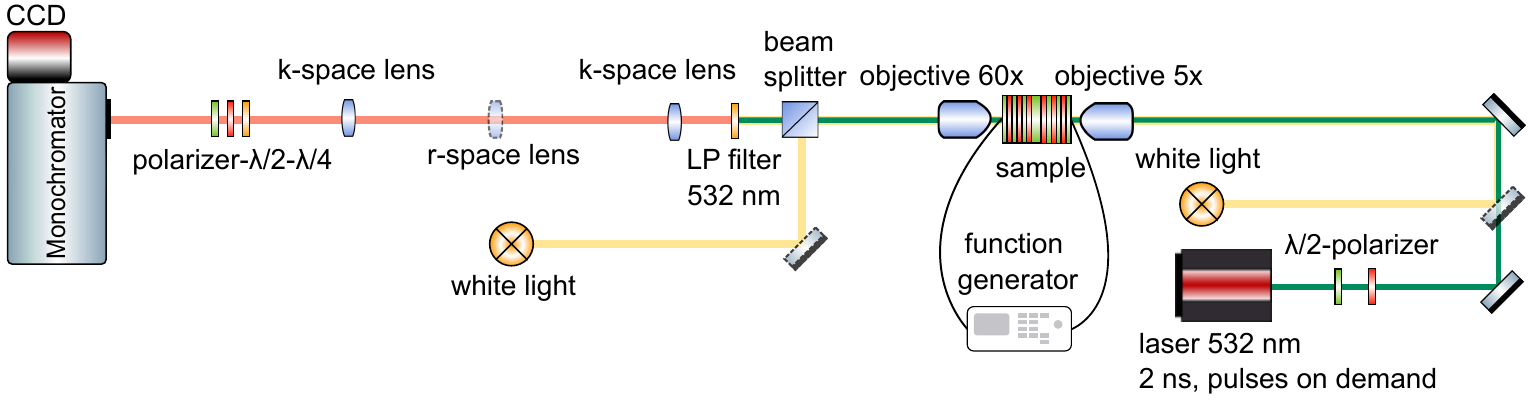}
    \caption{Scheme of the experimental optical setup.} 
    \label{im:S1}
\end{figure*}

When the voltage is applied to the cavity along the $z$ axis, molecules aligned closely to the electric field direction are the first to react on the field, and they realign with it even better, while molecules perpendicular to the field almost do not realign. This effect is often referred to as the dielectric effect of the electric field~\cite{rudquist1994linear}]. Thus, the uniformization of the director distribution along $z$ happens for some regions (edges of the distribution in Fig.~2b,d of the main text, $y/p \approx 0$ or $1/2$), while it does not happen for the others (middle of the distribution in Fig.~2b,d of the main text, $y/p \approx 1/4$). Therefore, the effective refractive index contrast is increasing for both polarizations: the highest refractive indices for $H$ and $V$ stay the same, while the lowest ones are decreasing, as shown in Fig.~\ref{im:sample_a}b. Also, the narrowing of the central region (high refractive index region) happens (more red-colored directors appear in Fig.~2b of the main text with respect to Fig.~2a of the main text). In complete analogy with zero voltage case, the potential is calculated by Eq.~\eqref{refr_index_to_potential} and shown in Fig.~\ref{im:sample_a}d. Now, the potential appears for $V$ polarization, with a localized state inside. For $H$ polarization, the potential deepens and allows now for two localized states. The continuum of states for both $H$ and $V$ is also shifted towards higher energy (shown by the pink and green colored rectangles, respectively). The described qualitative behavior agrees perfectly with the experimental observations presented in Fig.~2e-j of the main paper.

\subsection{Sample B}

Although it is well known that the helix of CLC creates the OA in an unbounded medium~\cite{de1993physics}, in our case, the OA is due to the combination of molecular tilt and helix, and not due to the helix alone.  
This can be explained from the point of view of system symmetry. The unbounded medium with a cholesteric helix does not show the inversion symmetry. This leads to the emergence of OA. When a cholesteric helix is bounded by the cavity in the ULH configuration, it can be unfolded to pave the space by building the reflection of the helix in the mirrors. In this case, the inversion symmetry (up to a supplementary translation by cavity width) is restored, preventing the appearance of OA. That is why the helix-induced OA does not show up in sample A (see Fig.~2 of the main paper). Finally, when a tilt is introduced, the inversion symmetry is broken again (along $x$), and OA is restored. That is what happens in sample B giving rise to the $s$-$p$ band hybridization.

\section{Experiment}

Figure~\ref{im:S1} presents the optical experimental setup described in detail in the Methods section of the main text. Depending on the selected lenses ($k$-space or $r$-space) and the light source, the setup allows measuring the reciprocal space (reflectivity of white light and photoluminescence), as well as the real space (transmission of white light) at the same position on the sample. Paths of white light are indicated by yellow lines, laser light paths are marked with green lines, and the signal from the sample is represented by an orange line.

\begin{figure*}
    \centering 
    \includegraphics[width=14 cm]{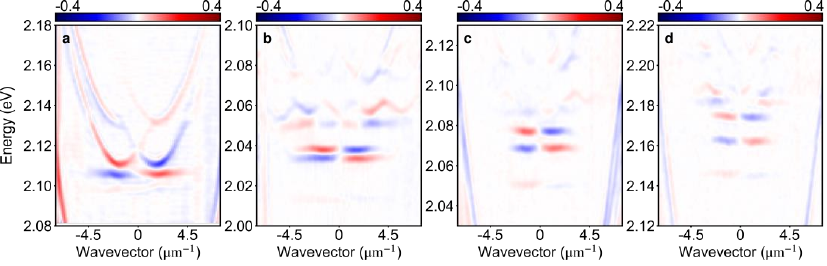}
    \caption{Degree of circular polarization expressed by Stokes parameter S$_3$ for different detunings between ISOC-mixed states. (a-d) Angle-resolved reflectivity spectra measured for different positions on sample B.} 
    \label{im:S2}
\end{figure*}

Fig.~\ref{im:S2} shows the Stokes parameter S$_3$ for angular-resolved reflectivity spectra measured at different positions on sample B. The positions were selected to have different detunings between the mixed ISOC states. It can be seen that with increasing detuning, the degree of circular polarization of mixed states decreases.

\begin{figure}
    \centering 
    \includegraphics[width=9 cm]{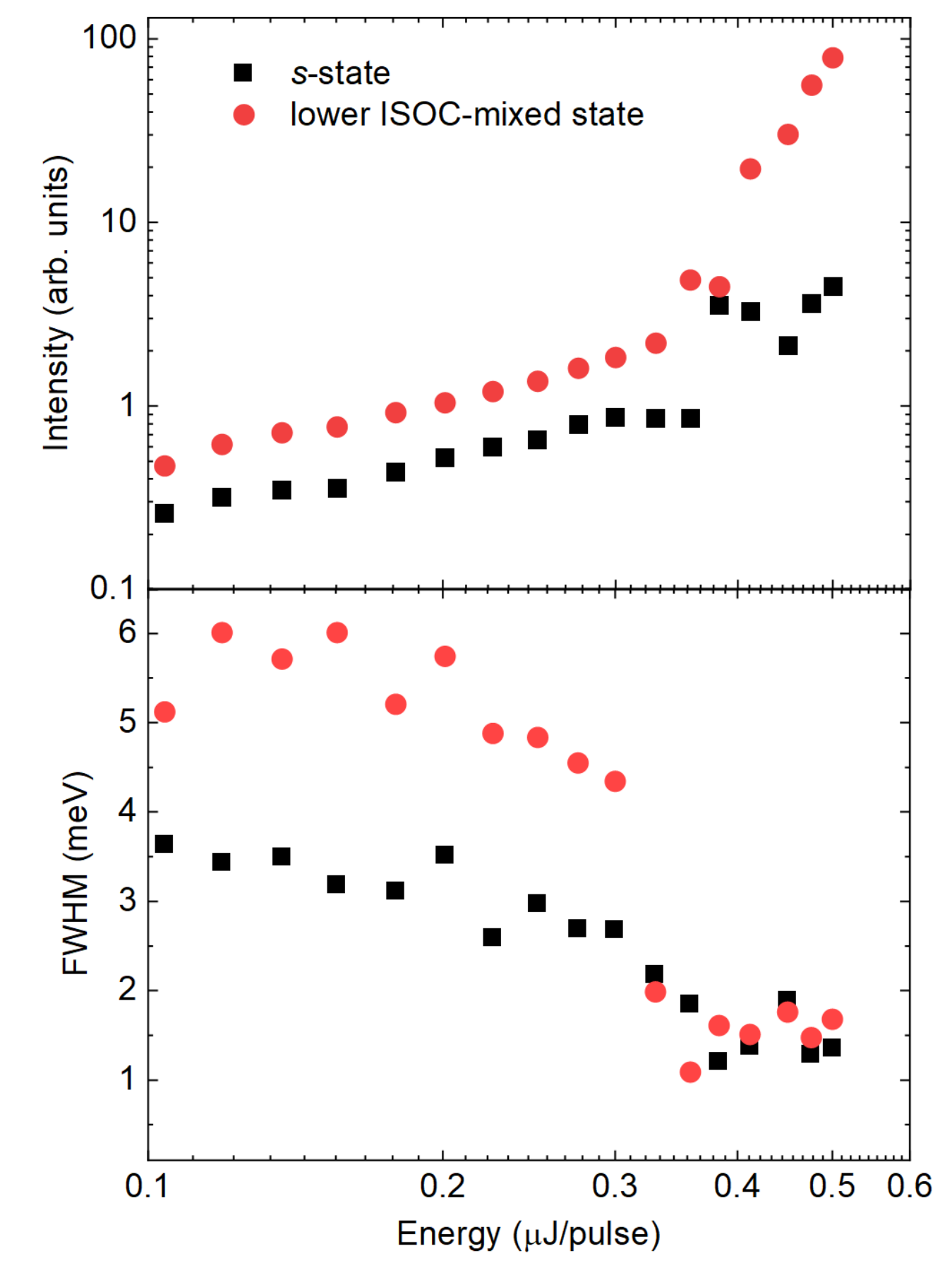}
    \caption{(a) Intensity and (b) full width at half maximum as a function of pumping pulse energy for horizontally polarized $s$-state and ISOC-mixed state.} 
    \label{im:S3}
\end{figure} 

Fig.~\ref{im:S3} shows the emission intensity (panel a) and full width at half maximum (panel b) of the two lowest ULH states ($H_s$ and the ISOC-mixed state) as a function of the excitation laser pulse energy measured for sample B. The values were obtained by fitting with a Gaussian curve the polarization- and momentum-integrated photoluminescence spectra. For a pulse energy of 0.36 {\textmu}J, a clear lasing threshold is observed, above which the emission intensity of both states exhibits an abrupt increase and the widths of both spectral lines undergo  a significant narrowing. Competition between the two lasing states is also present, where, above the threshold, the emission intensities fluctuate.

\section{Solution of Maxwell's equations based on a finite element method}

To get a deeper insight into physical mechanisms governing the observed effects, a series of numerical simulations was performed using COMSOL Multiphysics\textsuperscript{\textregistered} software. The 2D ($yz$) simulations were based on the Electromagnetic Waves, Frequency Domain Physics Interface of the Wave Optics Module and the Eigenfrequency Study. The computational domain is represented by a unit cell consisting of: (i) the microcavity domain of height $a_0 = 2$~$\mu$m corresponding to the ULH period and width $t_0 = 200$~nm corresponding to the effective microcavity thickness; (ii) the out-of-microcavity domains of width $s_0=160$~nm each. The Layered Transition Boundary Condition is applied to the boundaries separating the regions inside and outside the microcavity; the number of layers is set to $11$, where the first one has refractive index $n_{\mathrm{TiO}_2} = 2.436$, the second one has $n_{\mathrm{SiO}_2} = 1.456$, and the subsequent layers alternate. Their thicknesses $t_{\mathrm{TiO}_2/\mathrm{SiO}_2}$ are tuned so $t_{\mathrm{TiO}_2/\mathrm{SiO}_2} = \lambda_0/(4 n_{\mathrm{TiO}_2/\mathrm{SiO}_2})$, where $\lambda_0 = 550$~nm. The Floquet Periodic Boundary Condition is applied to the top and the bottom boundaries of the computational domain, whereas the Scattering Boundary Condition was applied to the side boundaries. The Scattering Boundary Condition is automatically tuned to the frequency searched around in the Eigenfrequency Study, which was set to $5 \times 10^{14}$~Hz ($\approx 2.07$~eV). Due to the system’s rectangular symmetry, the Mapped Mesh was used everywhere. 

The regions outside the microcavity have relative permittivity $\varepsilon=1$, relative permeability $\mu=1$ and conductivity $\sigma=0$. The region inside the microcavity has relative permeability $\mu=1$, conductivity $\sigma=0$ and relative permittivity defined as $\varepsilon_{\alpha,\beta} = n_{\mathrm{o}}^2 \, \delta_{\alpha,\beta} + (n_{\mathrm{e}}^2 - n_{\mathrm{o}}^2) \, \hat{r}_{\alpha} \hat{r}_{\beta}$, where $n_{\mathrm{o}} = 1.775$, $n_{\mathrm{e}} = 1.825$, $\hat{r}_{\alpha}, \hat{r}_{\beta}$ are the $\alpha = x,y,z$ and $\beta = x,y,z$ components of the liquid crystal director $\hat{\bm{r}}$, and $\delta_{\alpha,\beta}$ is the Kronecker symbol. The director $\hat{\bm{r}}$ is defined as

\begin{equation}
    \hat{\bm{r}} = \mathbf{R}_{x}(\alpha_x) \mathbf{R}_{y}(\alpha_y(y)) \hat{\bm{r}}^{(0)}\,,
\end{equation}
%
where

\begin{subequations}
    \begin{equation}
        \hat{\bm{r}}^{(0)} =
        \begin{bmatrix}
            0 \\
            0 \\
            1
        \end{bmatrix},
    \end{equation}
    \begin{equation}
        \mathbf{R}_{x}\left(\alpha_x \right) = 
        \begin{bmatrix}
            1 & 0 & 0 \\
            0 & \cos \left(\alpha_x \right) & -\sin \left(\alpha_x\right) \\
            0 & \sin \left(\alpha_x \right) & \cos \left(\alpha_x \right)
        \end{bmatrix},
    \end{equation}
    \begin{equation}
        \mathbf{R}_{y}(\alpha_y(y)) = 
        \begin{bmatrix}
            \cos \left(\alpha_y(y) \right) & 0 & \sin \left(\alpha_y(y) \right) \\
            0 & 1 & 0 \\
            -\sin \left(\alpha_y(y) \right) & 0 & \cos \left(\alpha_y(y) \right)
        \end{bmatrix}.
    \end{equation}
\end{subequations}

Here, the angle $\alpha_y(y) = \pi y/a_0$ is responsible for the ULH formation, whereas the angle $\alpha_x$ is responsible for the molecules’ tilt. 

\begin{figure*}[h]
    \centering
    \includegraphics{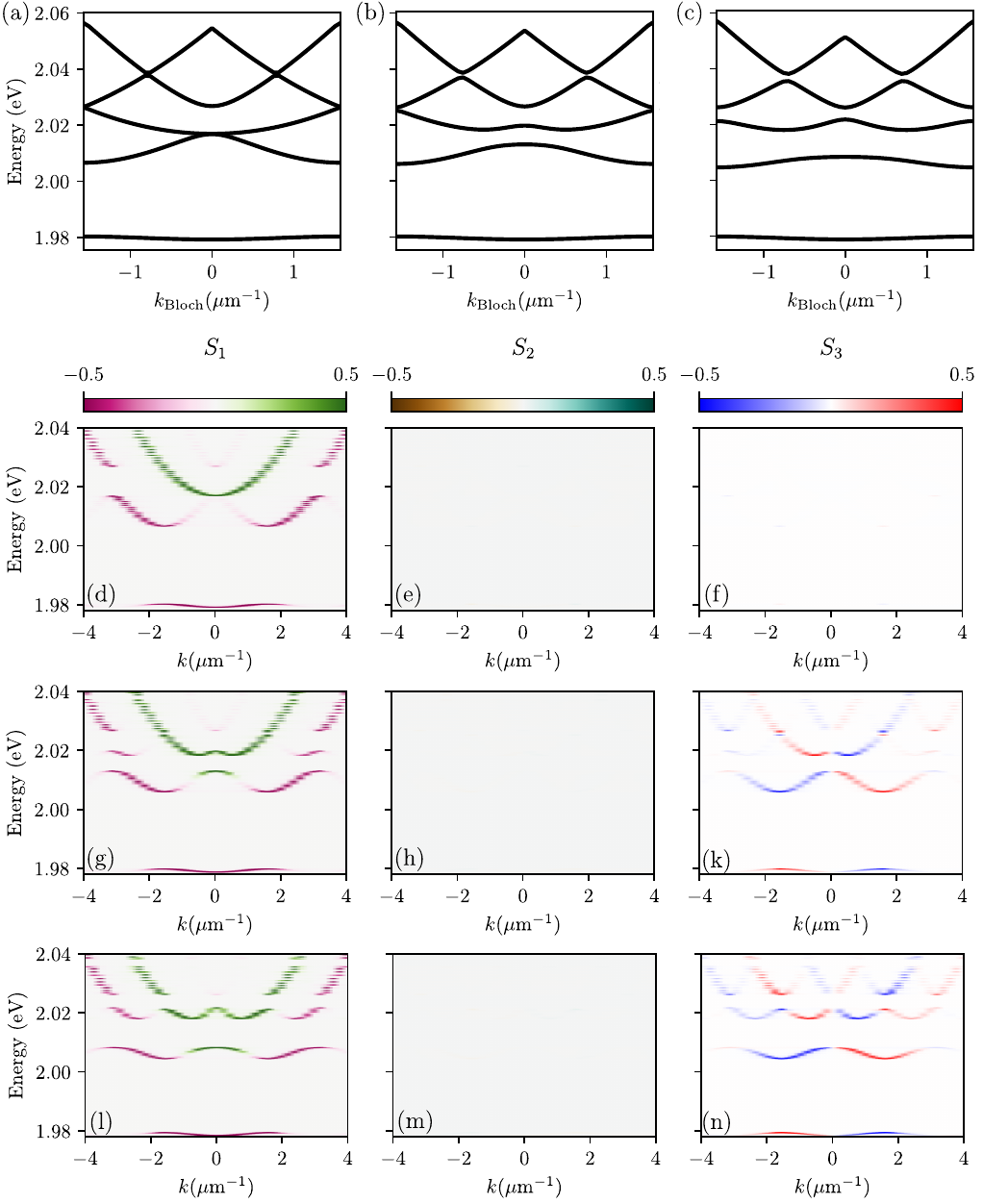}
    \caption{Simulation results. (a,b,c) Band structures for $\alpha_x = 0^{\circ}, 10^{\circ}, 20^{\circ}$, respectively. (d-f), (g-k), (l-n) $\mathrm{S}_{1-3}$ Stokes parameter corresponding to the first $5$ eigenenergies in reciprocal space for $\alpha_z = 0^{\circ}, 10^{\circ}, 20^{\circ}$, respectively.}
    \label{fig:simulations}
\end{figure*}

The obtained simulation results are represented by spatial distributions of the electric field Cartesian components $E_{x,y,z}$ corresponding to the found $32$ eigenfrequencies for each of $51$ considered Bloch wavenumbers. For the first five modes of interest derived by filtering the eigenfrequencies, periodic parts of $E_x$ and $E_y$ were calculated at $2^6$ points along a line parallel to the microcavity mirrors crossing the point $(y, z) = (0.5(t_0+s_0),0)$. These periodic parts were duplicated $2^4$ times and multiplied by the corresponding Bloch phase; then, the DFT of the obtained sequence of complex numbers was performed. Based on the Fourier images of the extended $E_{x,y}$ the Stokes parameters were calculated in reciprocal space. The results of post-processing for $\alpha_x = 0^{\circ}, 10^{\circ}, 20^{\circ}$ are presented in Fig.~\ref{fig:simulations} proving the necessity of tilt to observe optical activity. It is noteworthy that another tilt angle $\alpha_z$ (rotation around $z$) does not give optical activity along $k_y$.

\section{Tight-binding model}

In the simplest approximation, the potential created by ULH (Figs.~2a and 3a of the main text) for each polarization can be described as a cosine function: $~{U_{H,V}(y) = 0.5 A_{H,V} \cos \left(2qy\right)}$, where $A_{H,V}$ are the potential amplitudes for $H$ and $V$, $q = 2\pi/p$, and $p$ is the pitch of the ULH. When the tilt is present in a stable molecular director distribution, these linear polarizations are coupled.
 
This coupling can be well described by introducing a SOC between linear polarizations proportional to the wavevector $k_y$, which reads in ($H$,$V$) basis as $-2 i \alpha \sigma_x \partial_y$, with the SOC amplitude $\alpha$ and the first Pauli matrix $\sigma_x$. This SOC is equivalent to the OA in the regime of small in-plane wavevectors $k_y$ with respect to $q/\Delta n$. As OA, it appears due to the inversion symmetry breaking in the system. The SOC amplitude $\alpha$ is an energy-dependent function in general, but since we are aiming to reproduce only $H_p$ and $V_s$ band coupling, we can consider it to be a constant. Then, the Hamiltonian of the tilted ULH, including SOC (OA), can be written in ($H$,$V$) basis in real space as:
\begin{widetext}
\begin{equation}
    H_{cont} =
    \begin{pmatrix}
        -\frac{\hbar^2}{2m} \frac{\partial^2}{\partial y^2} + \frac{A_H}{2} [1 - \cos{\left(2qy\right)}] - \frac{\delta}{2} & -2i\alpha \frac{\partial}{\partial y} \\
        -2i\alpha \frac{\partial}{\partial y} & -\frac{\hbar^2}{2m} \frac{\partial^2}{\partial y^2} + \frac{A_V}{2} [1 - \cos{\left(2qy\right)}] + \frac{\delta}{2}
    \end{pmatrix},   \label{SM_continuous_Hamiltonian}
\end{equation}
\end{widetext}

with $\hbar$ reduced Planck constant, $m$ mass of photon in LC microcavity, and $\delta$ detuning between H and V potentials.
Since we would like to reproduce the band and polarization structure of low-energy bands $H_p$ and $V_s$ well-isolated from other bands, we can reduce the Hamiltonian~(\ref{SM_continuous_Hamiltonian}) to two-band tight-binding Hamiltonian. The cosine potential of each polarization can be well-approximated by harmonic oscillator potential for lower energy states, which allows us to use $s$ and $p$ eigenstates of a harmonic oscillator as basis states $\ket{0(y),V}$ and $\ket{1(y),H}$, respectively. Thus, the tight-binding Hamiltonian of ULH, including ISOC, writes in $k$-space as:
\begin{widetext}

\begin{subequations}
    \begin{equation}
        H = 
        \begin{pmatrix}
            E_{H_p} - 2t_{pp} \cos{(k_y p/2)} & -i\alpha \left[ t_{sp,0} - 2 t_{sp,1} \cos{(k_y p/2)} \right] \\
            c.c. & E_{V_s} - 2t_{ss} \cos{(k_y p/2)}
        \end{pmatrix},
        \label{SM_tb_Hamiltonian}
    \end{equation}

    \begin{equation}
        E_{H_p} = \bra{1(y),H} \left( -\frac{\hbar^2}{2m} \frac{\partial^2}{\partial y^2} + \frac{A_H}{2} [1 - \cos{\left(2qy\right)}] - \frac{\delta}{2} \right) \ket{1(y),H},
    \end{equation}
    \begin{equation}
        E_{V_s} = \bra{0(y),V} \left( -\frac{\hbar^2}{2m} \frac{\partial^2}{\partial y^2} + \frac{A_V}{2} [1 - \cos{\left(2qy\right)}] + \frac{\delta}{2} \right) \ket{0(y),V},
    \end{equation}
    \begin{equation}
        t_{pp} = \bra{1(y+p/2),H} \left( -\frac{\hbar^2}{2m} \frac{\partial^2}{\partial y^2} + \frac{A_H}{2} [1 - \cos{\left(2qy\right)}] - \frac{\delta}{2} \right) \ket{1(y),H},
    \end{equation}
    \begin{equation}
        t_{ss} = \bra{0(y+p/2),V} \left( -\frac{\hbar^2}{2m} \frac{\partial^2}{\partial y^2} + \frac{A_V}{2} [1 - \cos{\left(2qy\right)}] + \frac{\delta}{2} \right) \ket{0(y),V},
    \end{equation}
    \begin{equation}
        t_{sp,0} = \bra{1(y),H} 2 \partial_y \ket{0(y),V},
    \end{equation}
    \begin{equation}
        t_{sp,1} = \bra{1(y+p/2),H} 2 \partial_y \ket{0(y),V},
    \end{equation}
\end{subequations}
\end{widetext}
with $E_{H_p}$ and $E_{V_s}$ onsite energies of $H_p$ and $V_s$, respectively, $t_{pp}$ and $t_{ss}$ nearest-neighbour couplings between $p$ states of $H$ and $s$ states of $V$, respectively, $t_{sp,0}$ and $t_{sp,1}$ onsite and nearest-neighbour couplings between $H_p$ and $V_s$, respectively. The ISOC terms $t_{sp,0}$ and $t_{sp,1}$ can only couple modes of different parity ($s$ and $p$, for example) since the SOC (OA) is an odd function of $k_y$. The ISOC onsite coupling $\alpha t_{sp,0}$ is dominant over other coupling terms ($t_{pp},\ t_{ss},\ \alpha t_{sp,1}$), which is confirmed by efficient mixing of polarizations (high $S_3$ values in Fig.~3i of the main text), and large band splitting with respect to the band size. The parameters used for Fig.~3b,c of the main text: $E_{H_p}=24.24$ meV, $E_{V_s}=23.27$ meV, $t_{pp}=0.12$ meV, $t_{ss}=0.10$ meV, $\alpha=1.1$ meV$\cdot \mu$m, $t_{sp,0}=1.71$ $\mu$m$^{-1}$, $t_{sp,1}=0.06$ $\mu$m$^{-1}$, $p=8$ $\mu$m.

\section{Solution of Maxwell's equations based on the Fourier decomposition of the Dielectric tensor }
\subsection{Dielectric tensor}
In our approach, the cavity consists of the two $\delta$-mirrors, which are separated by distance $L$, and the volume between them is filled by a birefringent material. The full description of the orientation of liquid crystal molecules as a function of position is very complex and  can be found by using a method based on the thermodynamic potentials \cite{Jia_(2021)_Crystals, Oswald_(2000)_Phys.Rep.}. However, the only crucial dependence in this case is in the $y$ direction (parallel to mirror) and other dependencies can be taken into account in the effective approach. As shown in Fig.~1 of the main text the refractive index periodically changes in the $y$-direction, so the $ij$ element of the dielectric tensor $\hat{\varepsilon}$ can be expressed by expanding it into Fourier series:
\begin{equation}
    \varepsilon_{ij} \left( y \right) = \frac{1}{2}\sum_{n = 0}^{\infty} \varepsilon_{ij, n} e^{4\pi i n y/ d} + \varepsilon_{ij, n}^{\ast} e^{-4\pi i n y/ d}
    \label{equation: definition of epsilon element}
\end{equation}
where $\varepsilon_{ij, n}$ is a $n$-th coefficient in Fourier series, $d$ is a pitch (double period) of structure, and $\varepsilon_{ij, -n} = \varepsilon_{ij, n} ^{\ast}$, because $\varepsilon_{ij}$ is a real function and is a symmetric matrix $\hat{\varepsilon} = \hat{\varepsilon}^{T}$, where superscript $T$ denotes the transposition of a matrix. Each element $\varepsilon_{ij, n}$ is an effective element, which means that it includes the impact of modulation of the liquid crystal molecular director as a function of position in the $z$ direction. The experimental data shows that the amplitudes of modulations along $y$ are small compared to the refractive indices of the liquid crystal mixture, so it can be treated as a perturbation in calculations. It means that we can split our solution into two parts. In the first one, we found the expression for the unperturbed case -- we took into account only the elements for $n = 0$ from Eq.~\eqref{equation: definition of epsilon element}. In the second part, we find the expression for elements of the interaction matrix for all $n$ of $\hat{\varepsilon}$ matrix Eq.~\eqref{equation: definition of epsilon element}. 

\subsection{Linear operator}
In our approach, the elements of the $\hat{\varepsilon}$ matrix depend on the $y$ coordinate, so the band structure can occur only in this direction. But, since the $k_{x}$ component is crucial to obtain few polarization patterns, we take into account all elements of the wave vector in the plane of the cavity. In addition, the modulation along $y$ is small, so in the zero-order approximation we assume that the electric field in the $x-y$ plane has the same form as in \cite{Oliwa_(2024)_Phys.rev.res.}, so similarly to this work we define a second-order differential equation for a two-dimensional vector of the electric field:
\begin{multline}
    \hat{\mathcal{L}}\left(z; \omega, \bm{k}\right) \bm{E} = \\ =\left[ \left(\partial_{z} - \mathbf{A}^{\mathrm{T}} \right) \left(\partial_{z} - \mathbf{A} \right) - \mathbf{B} + k_{0}^{2} \mathbf{B}_{0} \right] \bm{E} = \bm{0}
\end{multline}
where $k_{0} = \omega/c$ and all matrices are equal to \mbox{$\mathbf{A} = \mathbf{A}_{x} + \mathbf{A}_{y}$}, $\mathbf{B} = \mathbf{B}_{x} + \mathbf{B}_{y} + \mathbf{B}_{xy}$, and:
\begin{subequations}
    \begin{equation}
        \mathbf{A}_{y} = - \frac{ik_{y}}{\varepsilon_{zz}}
        \begin{bmatrix}
            0 & 0 \\
            \varepsilon_{xz} & \varepsilon_{yz}
        \end{bmatrix}
    \end{equation}
    \begin{equation}
        \mathbf{A}_{x} = - \frac{i k_{x}}{\varepsilon_{zz}}
        \begin{bmatrix}
            \varepsilon_{xz} & \varepsilon_{yz} \\ 
            0 & 0 
        \end{bmatrix}
    \end{equation}
    \begin{equation}
        \mathbf{B}_{y} = \frac{k_{y}^{2}}{\varepsilon_{zz}}
        \begin{bmatrix}
            \varepsilon_{zz} & 0 \\ 
            \tilde{\varepsilon}_{yx} & \tilde{\varepsilon}_{yy}
        \end{bmatrix}
    \end{equation}
    \begin{equation}
        \mathbf{B}_{x} = \frac{k_{x}^{2}}{\varepsilon_{zz}}
        \begin{bmatrix}
            \tilde{\varepsilon}_{xx} & \tilde{\varepsilon}_{xy}\\
            0 & \varepsilon_{zz}
        \end{bmatrix}
    \end{equation}
    \begin{equation}
        \mathbf{B}_{xy} = \frac{k_{x}k_{y}}{\varepsilon_{zz}}
        \begin{bmatrix}
            \tilde{\varepsilon}_{yx} & \tilde{\varepsilon}_{yy} - \varepsilon_{zz} \\
            \tilde{\varepsilon}_{xx} - \varepsilon_{zz} & \tilde{\varepsilon}_{xy}
        \end{bmatrix}
    \end{equation}
    \begin{equation}
        \mathbf{B}_{0} = 
        \begin{bmatrix}
            \tilde{\varepsilon}_{xx} & \tilde{\varepsilon}_{xy} \\
            \tilde{\varepsilon}_{xy} & \tilde{\varepsilon}_{yy}
        \end{bmatrix}
    \end{equation}
\end{subequations}
where $\tilde{\varepsilon}_{ij} = \varepsilon_{ij} - \varepsilon_{iz} \varepsilon_{zj} / \varepsilon_{zz}$. The $\mathbf{B}_{0}$ matrix can be split into two parts: $\mathbf{B}_{0}^{0}$ which does not depend on the position and $\mathbf{B}_{0}^{1}$ which depends on position, so:
\begin{subequations}
    \begin{equation}
        \mathbf{B}_{0}^{0} = 
        \begin{bmatrix}
            \varepsilon_{xx, 0} & 0 \\
            0 & \varepsilon_{yy, 0}
        \end{bmatrix}
    \end{equation}
    \begin{equation}
        \mathbf{B}_{0}^{1} = \mathbf{B}_{0} - \mathbf{B}_{0}^{0}
    \end{equation}
\end{subequations}
Experimental data shows that the $\varepsilon_{xy, 0} = \varepsilon_{yx, 0} = 0$, so we simplify the formula for matrix $\mathbf{B}_{0}^{0}$, which yields two separate equations for two orthogonal linear polarizations for the unperturbed case. If these elements are different from zero, before further calculation we find the eigenvectors of matrix $\mathbf{B}$ and rotate the basis by a canonical transformation. In addition, $\varepsilon_{xz, 0} = 0$, but $\varepsilon_{yz} \neq 0$ so $\tilde{\varepsilon}_{xx, 0} = \varepsilon_{xx, 0}$ and $\tilde{\varepsilon}_{yy, 0} = \varepsilon_{yy, 0} - \varepsilon_{yz, 0}^{2}/\varepsilon_{zz, 0}$.
\subsection{Hamiltonian matrix}
According to \cite{Oliwa_(2024)_Phys.rev.res.} the general expression for the resonant state is given by:
\begin{subequations}
    \begin{equation}
        \hat{\mathcal{L}}\left( z, \omega_{r}, \bm{k} \right)\bm{u}_{r} = \bm{0},
        \label{equation: general equation for u state}
    \end{equation}
    \begin{equation}
        \hat{\mathcal{L}}^{\dag}\left( z, \omega_{r}, \bm{k} \right)\bm{w}_{r} = \bm{0},
        \label{equation: general equation for w state}
    \end{equation}
\end{subequations}
where $\omega_{r}$ is a resonant frequency and $\bm{u}_{r}$, $\bm{w}_{r}$ are resonant states for equation and coupled equation, and they are normalized, according to:
\begin{multline}
    \frac{1}{c^{2}}\int_{a}^{b} \bm{w}_{r}^{T} \left(z \right) \mathbf{B}_{0}^{0} \left( z \right) \bm{u}_{r} \left( z \right) \mathrm{d} z + \frac{i n_{a}^{2}}{2 c^{2} k_{zr}} \cdot \\ \cdot \left( \bm{w}_{r}^{T} \left( b \right) + \bm{u}_{r} \left( b \right) + \bm{w}_{r}^{T} \left( a \right) \bm{u}_{r} \left( a \right) \right) = 1
    \label{equation: normalization of state}
\end{multline}
where $\omega_{r}^{2} \left(\bm{k}\right)=  \left(k_{zr}^{2} + k_{x}^{2} + k_{y}^{2} \right) c^{2}/n_{a}^{2}$ and $k_{zr}$ is a resonant wave vector. Using the $\bm{u}_{r}$, $\bm{w}_{r}$ resonant state, we define the principal part of Green's function:
\begin{equation}
    \mathbf{G} \left(z, z^{\prime}; \omega, \bm{k} \right) \approx \frac{\bm{u}_{r} \left( z \right) \bm{w}_{r}^{T} \left(z ^{\prime} \right)}{2 \omega_{r} \left(\omega - \omega_{r}\right)}
\end{equation}
Solving Eq.~\eqref{equation: general equation for u state} in a general case is very difficult, but it can be done using perturbation theory. Similarly to \cite{Oliwa_(2024)_Phys.rev.res.}, we split the linear operator $\hat{\mathcal{L}}\left( z, \omega_{r}, \bm{k} \right)$ into resonant (unperturbed) $\hat{\mathcal{L}}_{0}$ and perturbed $\hat{\mathcal{L}}^{\prime}$ parts, where the first has the following form:
\begin{equation}
    \hat{\mathcal{L}}_{0} \left( z, \omega, \bm{k} \right) = \partial_{z}^{2} + \left( \frac{\omega^{2}}{c^{2}} - \frac{k_{x}^{2} + k_{y}^{2}}{n_{a}^{2}} \right) \mathbf{B}_{0}^{0} \left( z \right) 
\end{equation}
with unperturbed resonant solution $\overset{o}{\bm{u}}_{r}$ obtained from equation:
\begin{equation}
    \hat{\mathcal{L}}_{0} \left( z, \overset{o}{\omega}_{r}\left( 0 \right), 0 \right)\overset{o}{\bm{u}}_{r} = \bm{0}
    \label{equation: equation for unperturbed state}
\end{equation}
where $\overset{o}{\omega}_{r}^{2}\left( \bm{k} \right) = \overset{o}{\omega}_{r}^{2} \left( 0 \right) + c^{2} \left(k_{x}^{2} + k_{y}^{2} \right)/n_{a}^{2}$. The perturbed parts depend on $\omega$ and are different from zero for the region between mirrors:
\begin{multline}
    \hat{\mathcal{L}}^{\prime} = \hat{\mathcal{L}} - \hat{\mathcal{L}}_{0}   = -\mathbf{A}^{T} \partial_{z} - \partial_{z} \mathbf{A} + \\+ \mathbf{A}^{T} \mathbf{A} - \mathbf{B} + \frac{k_{x}^{2} + k_{y}^{2}}{n_{a}^{2}} \mathbf{B}_{0}^{0} + k_{0}^{2} \mathbf{B}_{0}^{1}
    \label{equation: L prime operatror}
\end{multline}
The resonant solution for $\bm{k} \neq \bm{0}$ can be expressed as a linear combination of unperturbed resonant state $\overset{o}{\bm{u}}_{n}$:
\begin{equation}
    \bm{u}_{m} \left( z \right) = \sum_{n} \frac{a_{nm}}{\sqrt{\overset{o}{\omega}_{n} \left( \bm{k} \right) }} \overset{o}{\bm{u}}_{r} \left( z \right)
\end{equation}
We used Mittag-Leffler theory and obtained expression for unperturbed $\mathbf{G}_{0} \left( z, z^{\prime}; \omega, \bm{k} \right)$ and perturbed $\mathbf{G} \left( z, z^{\prime}; \omega, \bm{k} \right)$ Green's functions:
\begin{subequations}
    \label{equation: perturbed and uperturbed Green's functions} 
    \begin{equation}
        \mathbf{G}_{0} \left( z, z^{\prime}; \omega, \bm{k} \right) = {\sum_{m}}^{\prime} \frac{\overset{o}{\bm{u}}_{m} \left( z \right) \overset{o}{\bm{u}}_{m}^{T} \left( z^{\prime} \right)}{2 \overset{o}{\omega}_{m} \left( \bm{k} \right) \left( \omega - \overset{o}{\omega}_{m} \left( \bm{k} \right) \right)},
    \end{equation}
    \begin{equation}
        \mathbf{G} \left( z, z^{\prime}; \omega, \bm{k} \right) = {\sum_{m}}^{\prime} \frac{\overset{o}{\bm{u}}_{m} \left( z \right) \overset{o}{\bm{w}}_{m}^{T} \left( z^{\prime} \right)}{2 \overset{o}{\omega}_{m} \left( \bm{k} \right) \left( \omega - \overset{o}{\omega}_{m} \left( \bm{k}\right) \right)}, 
    \end{equation}
\end{subequations}
where the summation is over simple poles associated with the resonant state limited to the poles in the lower half of the complex $\omega$ plane, including the pole at zero if exists. To determine the elements of the Hamiltonian matrix, we consider the Dyson formula for $\mathbf{G}\left( z, z^{\prime}; \omega, \bm{k} \right)$:
\begin{multline}
    \mathbf{G} \left( z, z^{\prime} \right) = \mathbf{G}_{0} \left( z, z^{\prime} \right) - \\ +\int_{a}^{b} \mathbf{G}_{0} \left( z, z^{\prime \prime} \right)  \hat{\mathbf{\mathcal{L}}}^{\prime} \left( z^{\prime \prime} \right) \mathbf{G} \left( z^{\prime \prime}, z^{\prime} \right) \mathrm{d} z^{\prime \prime}  
    \label{equation: Dyson formula}
\end{multline}
Inserting Eq.~\eqref{equation: L prime operatror} and Eq.~\eqref{equation: perturbed and uperturbed Green's functions} to Eq.~\eqref{equation: Dyson formula}, after transformation presented in \cite{Oliwa_(2024)_Phys.rev.res.} yields a general eigenvalue problem:
\begin{equation}
    \sum_{p} \mathcal{H}_{m p} \left( \bm{k} \right) a_{p} = \omega \left( \bm{k} \right) a_{m}
\end{equation}
with  a non-Hermitian Hamiltonian:
\begin{equation}
    \mathcal{H}_{mp} = \overset{o}{\omega}_{m} \left( \bm{k} \right) \delta_{m p} - \frac{ \hat{\bm{\mathcal{L}}}^{\prime}_{mp}}{2 \sqrt{ \overset{o}{\omega}_{m} \left( \bm{k} \right) \overset{o}{\omega}_{p} \left( \bm{k} \right)}}
\end{equation}
where
\begin{equation}
    \hat{\bm{\mathcal{L}}}^{\prime}_{mp} = \int_{a}^{b} \overset{o}{\bm{u}}^{T}_{m} \left( z^{\prime \prime} \right) \hat{\bm{\mathcal{L}}}^{\prime} \left( z ^{\prime \prime} \right) \overset{o}{\bm{u}}^{T}_{p} \left( z^{\prime \prime} \right) \mathrm{d} z^{\prime \prime}
\end{equation}

In the case under consideration, $\hat{\bm{\mathcal{L}}}^{\prime} (z)$ includes the term $k_{0}^{2} \mathbf{B}_{0}^{1}$ (see Eq.~\eqref{equation: L prime operatror}), necessitating the solution of a general eigenvalue problem. This scenario is referred to as a quadratic eigenvalue problem (QEP), which can be addressed by applying linearization theory. Within the analyzed energy range, $k_{0}^{2}$ varies only slightly, allowing us to approximate this term as having a constant value in the zeroth order of approximation.

\subsection{Electric field in simple cavity}
The coordinate dependence of refractive index in the system has the following form:
\begin{multline}
    \tilde{n}_{s}^{2} \left( z \right) = \Theta \left( L - z \right) \Theta \left( z \right) \left( n_{s}^{2} - n_{a}^{2} \right) + n_{a}^{2} -  \\ + \frac{2\zeta_{s}c^{2}}{\omega^{2}} \left( \delta \left( z \right) +  \delta \left( L- z \right) \right)
\end{multline}
where $n_{a}$ is a ambient refractive index outside cavity, $n_{s} = \sqrt{\tilde{\varepsilon}_{ss, 0}}$ denotes the refractive index for $s = H$ (horizontal mode) and $s = V$ (vertical mode), $\zeta_{s}$ is the strength of the $\delta$-mirrors and it depends on polarization, and $\Theta \left( z \right)$ is a standard Heaviside step function. In this case, the equation for the electric field for the perpendicular incident wave has the following form:
\begin{equation}
    \left( \partial_{z}^{2} + \tilde{n}_{s}^{2} \left( z \right) \frac{\omega^{2}}{c^{2}} \right) \bm{E}_{s} \left( z \right) = \bm{0}
\end{equation}
which is the same as the Eq.~\eqref{equation: equation for unperturbed state} and has following solution:
\begin{equation}
    \bm{E}_{sm} \left( z \right) = \bm{E}_{0, sm} \sin \left( \kappa_{sm} \left( z - d_{sm} \right) \right)
\end{equation}
with normalization accordingly to Eq.~\eqref{equation: normalization of state}, where $\kappa_{sm}$ is defined in the implicit equation:
\begin{subequations}
    \begin{equation}
        \kappa_{sm} = -\frac{i}{L} \ln \left( \frac{1 - i \frac{ \left(n_{a} + n_{s} \right) \kappa_{sm} } {2 \zeta_{s} n_{s}}} {1 - i \frac{ \left( n_{a} - n_{s} \right) \kappa_{sm}} {2 \zeta_{s} n_{s}}} \right) + \frac{m \pi}{L} 
    \end{equation}
\end{subequations}
and $d_{sm} = \Delta_{sm}L/2\kappa_{sm}$, where:
\begin{multline}
    \Delta_{sm} = \kappa_{sm} - \frac{m\pi}{L} \approx \\ \approx -\frac{m \pi}{L^{2} \zeta_{s}} \left[ 1 - \frac{1}{L \zeta_{s}} \left(1 - i \frac{n_{a} m \pi}{2n_{s}} \right) \right] 
\end{multline}
The $d_{sm}$ parameter is a scattering length and for ideal cavity $\zeta_{s} \rightarrow \pm \infty$ goes to 0 because $\Delta_{sm} \rightarrow 0$, where $\Delta_{sm}$ denotes the difference between the ideal and non-ideal Fabry-Perot resonator. The $\zeta_{s}$ parameter is crucial in this approach as it defines the photon lifetime in the cavity and significantly influences the broadening of modes. 
\subsection{Hamiltonian}
According to \cite{Oliwa_(2024)_Phys.rev.res.}, we can define matrices $\mathbf{A}_{1} = \mathbf{A} + \mathbf{A}^{T}$, $\mathbf{B}_{1} = \mathbf{B} - \mathbf{A}^{T} \mathbf{A}$ and constant \mbox{$\chi_{sm, s^{\prime} m^{\prime}} = c^{2}/2n_{s}n_{s^{\prime}} \sqrt{\overset{o}{\omega}_{sm}\left( 0 \right) \overset{o}{\omega}_{s^{\prime} m^{\prime}} \left( 0 \right)}$}. In this case, the approximated element of the multimode Hamiltonian is equal to:
\begin{multline}
    \mathcal{H}_{sm, s^{\prime}m^{\prime}} = \delta_{ss^{\prime}}\delta_{m m^{\prime}}\overset{o}{\omega}_{sm} \left(\bm{k}\right) +  \\ + \chi_{sm, s^{\prime}m^{\prime}} \left(\mathbf{A}_{1}\right)_{s s^{\prime}} Q_{sm, s^{\prime}m^{\prime}} +  \\ + \chi_{sm, s^{\prime}m^{\prime}} \left(\mathbf{B}_{1} - \frac{k_{x}^{2} + k_{y}^{2}}{n_{a}^{2}} \mathbf{B}_{0}^{0} - k_{0}^{2} \mathbf{B}_{0}^{1} \right)_{ss^{\prime}} P_{sm, s^{\prime}m^{\prime}}
    \label{equation: multimode hamiltonian}
\end{multline}
where $P_{sm, s^{\prime} m^{\prime}}$ and $Q_{sm, s^{\prime} m^{\prime}}$ are equal to:
\begin{subequations}
    \begin{multline}
        P_{sm,s'm'} = \frac{n_sn_{s'}}{c^2}\int_0^L \left(\bm{E}_{sm}\left(z \right) \right)^T \bm {E}_{s'm'} \left(z\right)\, \mathrm{d}z
        \approx\\  \approx 
        \begin{cases}
            0 & \text{if $m$+$m'$ odd,} \\
            1 & \text{if $m$ = $m'$,} \\
            \frac{2L(m'\Delta_{sm} - m\Delta_{s'm'})} {\pi (m^2-m'^2)} & \text{in other cases} \\
        \end{cases}
    \end{multline}
    and
    \begin{multline}
        Q_{sm,s'm'} = \frac{n_sn_{s'}}{c^2}\int_0^L \left(\bm E_{sm}\left(z \right) \right)^T\partial_z \bm {E}_{s'm'} \left(z \right)\, \mathrm{d}z \approx\\ 
        \approx 
        \begin{cases}
            0 & \text{if $m$+$m'$ even,} \\
            \frac{4mm'}{L (m^2-m'^2)} + \frac{8(m^3\Delta_{s'm'} - {m'}^3\Delta_{sm})}{\pi (m^2-m'^2)^2} &  \text{if $m$+$m'$ odd} \\
        \end{cases}
    \end{multline}
\end{subequations}
In this approach, we assumed that the modulation in the $z$ direction is minimal, allowing us to treat each element of the matrices $\mathbf{A}_{1}$, $\mathbf{B}_{1}$, $\mathbf{B}_{0}^{0}$ and $\mathbf{B}_{0}^{1}$  as independent of the $z$ coordinate. Consequently, these elements can be extracted from the integral and will be considered in further calculations as an 'effective value.' We postulate that each amplitude in the Fourier series for elements of the dielectric tensor, as defined in Eq.~\eqref{equation: definition of epsilon element}, acts as a fitting parameter. Thus, we incorporate the $z$-direction dependence by adjusting the $z$-averaged effective amplitude values.

In the subsequent step, we simplify the multimode Hamiltonian to a 2-mode Hamiltonian, capturing the interaction between the nearest modes within the measurement range. Experimental data indicates the presence of a coupling between cavity modes of the same mode number, leading to the following formula for the $2 \times 2$ Hamiltonian matrix:
\begin{equation}
    \mathcal{H}_{sm, s^{\prime} m} = \delta_{ss^{\prime}}\overset{o}{\omega}_{s m} + \chi_{sm, s^{\prime} m} \left(\mathbf{B}_{1} - k_{0}^{2}\mathbf{B}_{0}^{1} \right)_{s s^{\prime}}
    \label{equation: 2 mode Hamiltonian}
\end{equation}
where $\overset{o}{\omega}_{s m} = \kappa_{sm} c/n_{s}$. In the considered case, the component of the dielectric tensor depends on the position $y$, so in the 2-mode Hamiltonian there appears the product of an element of dielectric tensor and $k_{y} = -i \hbar \partial/\partial y$, so we have $k_{y}\varepsilon \left(y \right)$ and $k_{y}^{2}\varepsilon \left(y \right)$, which can be expressed in the symmetric form. The modulation $\hat{\varepsilon}$ in the $z$ direction is low, so the electric field can be expressed as $\bm{E}_{sm} \left(y, z\right) = \bm{E}_{sm} \left(z \right) E_{sm} \left( y \right)$, where $E_{sm} \left(y \right)$ is a normalized periodic function with period $d$ due to the periodicity of $\hat{\varepsilon}$. Using the above assumption we derived the formula for the element of the Hamiltonian matrix:
\begin{multline}
    \mathcal{H}_{sm, s^{\prime} m} = \delta_{ss^{\prime}}\overset{o}{\omega}_{s m} + \\ + \chi_{sm, s^{\prime} m^{\prime}} \left\langle E_{sm} \left(y\right) \left| \left(\mathbf{B}_{1}\right)_{s s^{\prime}} \right| E_{s^{\prime} m^{\prime}} \left( y \right) \right\rangle - \\ + k_{0}^{2}\chi_{sm, s^{\prime} m^{\prime }} \left \langle E_{sm} \left( y \right) \left| \left(\mathbf{B}_{0}^{1}\right)_{s s^{\prime}} \right| E_{s^{\prime}m^{\prime}} \left( y \right) \right \rangle 
\end{multline}
The first and third components are independent of the $k_{y}$ and $k_{y}^{2}$  operators, so we applied the symmetrization procedure solely to the second term. This results in the following form for the product of the derivative operator and the dielectric tensor:

\begin{subequations}
    \label{equation: symetrized operator}
    \begin{multline}
        \left\langle E_{sm} \left(y\right) \left| \varepsilon_{1, ss^{\prime}} \left(y\right) k_{y} \right| E_{s^{\prime} m^{\prime}} \left( y \right) \right\rangle = \\ = -\frac{i}{2} \int_{-d/2}^{d/2} 2 E_{s m}^{\ast} \left(y \right) \varepsilon_{1, s s^{\prime}} \left( y \right) \frac{\mathrm{d}E_{s^{\prime}m^{\prime}} \left( y \right)}{\mathrm{d} y} + \\ + E_{sm}^{\ast} \left(y \right) E_{s^{\prime} m^{\prime}} \left( y \right) \frac{\mathrm{d} \varepsilon_{1, ss^{\prime}} \left(y\right)}{\mathrm{d} y} \mathrm{d} y = \\ = \left\langle E_{sm} \left(y\right) \left| \varepsilon_{1, ss^{\prime}} \left(y\right) k_{y} \right| E_{s^{\prime} m^{\prime}} \left( y \right) \right\rangle ^{\ast},
    \end{multline}
    \begin{multline}
        \left\langle E_{sm} \left(y\right) \left| \varepsilon_{2, ss^{\prime}} \left(y\right) k_{y}^{2} \right| E_{s^{\prime} m^{\prime}} \left( y \right) \right\rangle = \\ = -\int_{-d/2}^{d/2} E_{s m}^{\ast} \left(y \right) \frac{\mathrm{d} \varepsilon_{2, s s^{\prime}} \left( y \right)}{\mathrm{d} y} \frac{\mathrm{d} E_{s^{\prime} m^{\prime}} \left( y \right)}{\mathrm{d} y} + \\ + E_{sm} ^{\ast} \left(y \right) \varepsilon_{2, s s^{\prime}} \left(y\right) \frac{\mathrm{d}^{2} E_{s^{\prime}m^{\prime}} \left( y \right)}{\mathrm{d} y^{2}} \mathrm{d} y = \\ = \left\langle E_{sm} \left( y \right) \left| \varepsilon_{2, ss^{\prime}} \left(y\right) k_{y}^{2} \right| E_{s^{\prime} m^{\prime}} \left( y \right) \right\rangle^{\ast}
    \end{multline}
\end{subequations}
where $\varepsilon_{1, s s^{\prime}} \left( y \right)$ and $\varepsilon_{2, s s^{\prime}} \left( y \right)$ denote the $s s^{\prime}$ element of matrix $\mathbf{B}_{1}$ which is proportional to $k_{y}$ and $k_{y}^{2}$, respectively. The hermiticity of operators in Eq.~\eqref{equation: symetrized operator} can be proved by using integration by parts because the $E_{sm} \left(y \right)$ and $\varepsilon_{ss^{\prime}} \left( y \right)$ are periodic function with period $d$.

\subsection{Dispersion relation}

Each element of the Hamiltonian matrix depends on the $y$ coordinate because $\hat{\varepsilon}$ varies with it. To derive energy and polarization as functions of position in real space or wave vector in reciprocal space, we discretize the $y$ coordinate. In this context, $k_{y} = -i\partial/\partial y$ and $k_{y}^{2} = - \partial^{2}/\partial y^{2}$ are substituted by finite difference approximations, with the finite difference coefficients calculated as per Fornberg (1988) \cite{Fornberg_(1988)_AMS}. Diagonalizing this Hamiltonian yields $2N$  eigenvalues $\Lambda$ (the set of all eigenvalues) and $2N$ eigenvectors $Q$ (the set of all eigenvectors), where $N$  represents the number of points in the lattice. Each eigenvector has $2N$  components, with the first $N$ components corresponding to the amplitude as a function of position for horizontally polarized modes, and the second $N$ for vertically polarized modes. The Stokes polarization parameters as functions of position and energy (in real space) were calculated using the following procedure:
\begin{subequations}
    \label{equation: Stokes polarization parameters}
    \begin{multline}
        S_{0, i j} = \sum_{m = 1}^{N} \sum_{n = 1}^{2N} \left(\left|Q_{m, n}\right|^{2} + \left|Q_{m+N, n}\right|^{2}\right) \cdot \\ 
        \cdot f_{1} \left(\mathrm{Re}\left(E_{n} - E_{i} \right), \mathrm{Im}\left(E_{n}\right) \right) f_{2} \left(y_{j} - y_{m}, \sigma \right),
    \end{multline}
    \begin{multline}
        S_{1, i j} = \sum_{m = 1}^{N} \sum_{n = 1}^{2p} \left(\left|Q_{m, n}\right|^{2} - \left|Q_{m + N, n}\right|^{2}\right) \cdot \\ 
        \cdot f_{1} \left(\mathrm{Re}\left(E_{n} - E_{i} \right), \mathrm{Im}\left(E_{n}\right) \right) f_{2} \left(y_{j} - y_{m}, \sigma \right),
    \end{multline}
    \begin{multline}
        S_{2, i j} = \sum_{m = 1}^{N} \sum_{n = 1}^{2p} \left(Q_{m + N, n}^{\ast}Q_{m, n} + Q_{m, n}^{\ast}Q_{m + N, n}\right) \cdot \\ 
        \cdot f_{1} \left(\mathrm{Re}\left(E_{n} - E_{i} \right), \mathrm{Im}\left(E_{n}\right) \right) f_{2} \left(y_{j} - y_{m}, \sigma \right),
    \end{multline}
    \begin{multline}
        S_{3, i j} = \sum_{m = 1}^{N} \sum_{n = 1}^{2p} i\left(Q_{m + N, n}^{\ast}Q_{m, n} - Q_{m,n}^{\ast}Q_{m + N, n} \right) \cdot \\ 
        \cdot f_{1} \left(\mathrm{Re}\left(E_{n} - E_{i} \right), \mathrm{Im}\left(E_{n}\right) \right) f_{2} \left(y_{j} - y_{m}, \sigma \right),
    \end{multline}
\end{subequations}
where  $S_{k, i, j}$ denote the value of $k$-th component of Stokes polarization parameters on the $i,j$ position on the grid and $f_{1}$, $f_{2}$ are the Lorentz and Gaussian distributions:
\begin{subequations}
    \begin{equation}
        f_{1} \left(x, \gamma \right) = \frac{1}{1 + x^{2}/\gamma^{2}}
    \end{equation}
    \begin{equation}
        f_{2} \left(x, \sigma \right) = \exp\left( -\frac{x^{2}}{\sigma^{2}} \right)
    \end{equation}
\end{subequations}
responsible for smoothing the solutions of discrete Hamiltonian. 

The amplitude as a function of the wave vector for horizontally and vertically polarized modes was determined using the discrete Fourier transform. Given that the Hamiltonian comprises two subsets of modes, the Fourier transform was performed separately for each subset:
\begin{equation}
    \tilde{Q}_{m, n} = \left\{ \begin{matrix} 
    \sum_{k = 1}^{N}Q_{k, n}e^{-2\pi i k m/N} & \mathrm{if} \, m \leq N \\
    \sum_{k = 1}^{N}Q_{k + N, n}e^{-2\pi i k m/N} & \mathrm{if} \, m > N 
    \end{matrix} \right.
\end{equation}
The Stokes polarization parameters as a function of energy and wave vector were calculated according to Eq.~\eqref{equation: Stokes polarization parameters}, where $Q$ and position $y$ were replaced with $\tilde{Q}$ and wave vector $k$, respectively.  
\subsection{Parameters in simulation}
\begin{table}[t]
    \centering
    \caption{Parameters used in the simulation.}
    \begin{tabular}{|c|c|c|c|} 
        \hline
        Parameter & value & Parameter & value \\ 
        \hline
        $L$ & $2.149 \,\mathrm{\mu m}$ & $d$ & $6.5 \,\mathrm{\mu m}$\\ 
        $m_{H}$ & 11 & $m_{V}$ & 11 \\
        $\zeta_{H}$ & $67\,\mathrm{\mu m^{-1}}$ & $\zeta_{V}$ & $65\,\mathrm{\mu m^{-1}}$ \\
        $\varepsilon_{xx, 0}$ & 2.500 & $\varepsilon_{yy, 0}$ & 2.472 \\
        $\varepsilon_{xx, 1}$ & 0.080 & $\varepsilon_{yy, 1}$ & 0.070 \\
        $\varepsilon_{xx, 2}$ & $-0.025$ & $\varepsilon_{yy, 2}$ & $-0.025$ \\
        $\varepsilon_{xx, 3}$ & 0.020 & $\varepsilon_{yy, 3}$ & $-0.015$ \\
        $\varepsilon_{xx, 4}$ & $-0.010$ & $\varepsilon_{yy, 4}$ & 0.010 \\
        $\varepsilon_{zz, 0}$ & 2.675 & $\varepsilon_{zz, 1}$ & $-0.150$ \\
        $\varepsilon_{zz, 2}$ & 0.050 & $\varepsilon_{zz, 3}$ & $-0.005$ \\ 
        $\varepsilon_{xy, 1}$ & $0.002 - 0.010i$ &
        $\varepsilon_{yz, 0}$ & 0.150 \\ $\varepsilon_{yz, 1}$ & $0.100$ & $\varepsilon_{xz, 1}$ & $0.080i$ \\ 
        $n_{e}$ & 1.737 & $n_{o}$ & 1.522 \\
        \hline
    \end{tabular}
    \label{table: value of parameter}
\end{table}

In our simulation's initial step, we estimate the cavity's thickness $L$, the number of modes ($m_{H}$ and $m_{V}$), and the linewidth $\zeta_{H}$ and $\zeta_{V}$ based on the modes' energy, as shown in Tab.~\ref{table: value of parameter}. Utilizing the nearly free electron model, we estimate the structure's period ($d$), consistent with real-space measurements. Next, we determine the amplitudes of $\varepsilon_{xx, n}$ and $\varepsilon_{yy, n}$ in Eq.~\eqref{equation: definition of epsilon element} to align each band's energy with the experimental data. The value of $\varepsilon_{zz, n}$ is selected to keep the $\hat{\varepsilon}$ matrix's trace constant in space, satisfying $\varepsilon_{zz, n} = - \varepsilon_{xx, n} - \varepsilon_{yy, n}$ for $n \neq 0$, and $\varepsilon_{zz, 0} = 2{n}_{o}^{2} + n_{e}^{2} - \varepsilon_{xx, 0} - \varepsilon_{yy, 0}$ for $n = 0$, where $n_{o}$ and $n_{e}$ are the ordinary and extraordinary refractive indices for the liquid crystal, estimated from the data for E7 liquid crystal mixture presented in \cite{Li_(2005)_JDT}. The value of $\varepsilon_{yz, 0}$ comes from comparing the experimentally obtained and theoretically predicted effective mass for the vertically-polarized mode. The values of $\varepsilon_{xy, 1}$, $\varepsilon_{xz, 1}$, and $\varepsilon_{yz, 1}$ are estimated by comparing the real and reciprocal space distribution of the Stokes polarization parameter $S_{2}$ and the reciprocal space's $S_{3}$ parameters. These sine components' values indicate the strength of interband spin-orbit coupling. All $\varepsilon_{ij, n}$ values not listed in Tab.~\ref{table: value of parameter} are set to zero. Additionally, we assume a slightly non-perpendicular incident wave, setting $k_{x} = 1\,\mathrm{\mu m^{-1}}$. The polarization patterns in Fig.~3g-l of the main paper were derived using Eq.~\eqref{equation: Stokes polarization parameters} on a grid matching the measurements. The grid for the Hamiltonian with periodic boundary conditions consists of $N = 2^{10}$ points, with the simulation window ranging from $-60 \,\mathrm{\mu m}$ to $60 \,\mathrm{\mu m}$. Additionally, we carried out calculations for the Hamiltonian that accounts for interactions with other states: the 10th, 11th, and 12th horizontally and vertically polarized modes, as detailed in Eq.~\eqref{equation: multimode hamiltonian}. The results indicate that the polarization patterns in reciprocal and real space closely resemble those derived from the 2-mode Hamiltonian. This confirms the validity of simplifying the multimode Hamiltonian to a 2-mode Hamiltonian.